\documentclass[a4paper,11pt]{article}
\pdfoutput=1 % if your are submitting a pdflatex (i.e. if you have
\usepackage{jcappub} % for details on the use of the package, please
\usepackage{amsmath,graphicx,hyperref,bm,hyperref}
\usepackage{color}
\title{\boldmath The cosmological vector modes from a monochromatic primordial power spectrum}

\author[a,b]{Zhe Chang}
\author[a,b]{Xukun Zhang,}
\author[a,b,1]{Jing-Zhi Zhou,\note{Corresponding author.}}

\affiliation[a]{Institute of High Energy Physics, Chinese Academy of Sciences, Beijing 100049, China}
\affiliation[b]{University of Chinese Academy of Sciences, Beijing 100049, China}

\emailAdd{changz@ihep.ac.cn}
\emailAdd{zhangxukun@ihep.ac.cn}
\emailAdd{zhoujingzhi@ihep.ac.cn}

\begin{abstract}
	{The cosmological background of higher order vector modes can be generated by the first order scalar perturbations. We investigate the second order and the third order vector modes systematically. The explicit expressions of two point functions $\langle V^{(n),\lambda}V^{(n),\lambda'} \rangle$$\left(n=2,3\right)$ and power spectra corresponded are presented. In the case of a monochromatic primordial power spectrum, the second order vector modes do not exist. However, the third order vector modes can be generated by a monochromatic primordial power spectrum. And it is found that the third order vector modes sourced by the second order scalar perturbations dominate the two point function $\langle V^{(3),\lambda}V^{(3),\lambda'} \rangle$ and power spectrum corresponded.}
\end{abstract}

\begin{document}
\maketitle
\flushbottom

\section{Introduction}
In the inflation theory, the cosmological perturbations are originated from the quantum fluctuations during inflation. The information about the early Universe is encoded in these perturbations. The cosmological perturbations can be decomposed as scalar, vector, and tensor perturbations based on symmetry of Friedmann-Robertson-Walker (FRW) spacetime. According to the observations of the cosmic microwave background (CMB) and large-scale structure, the primordial power spectra are well constricted on large scales ($\gtrsim$1 Mpc) \cite{Abdalla:2022yfr,Planck:2018vyg}. It indicates a nearly scale-invariant primordial power spectrum of scalar perturbations with amplitude $\backsim 2\times 10^{-9}$, and tensor perturbations with tensor-to-scalar ratio $r<0.06$. However, contrary to the scalar modes and the tensor modes, the vector modes decay as $1/a^2$ after they leave the Hubble horizon during inflation, and the scalar perturbations can not generate vector modes at linear order \cite{Bassett:2005xm}. Therefore, the vector perturbations are always neglected at first order.

On small scales ($\lesssim$1 Mpc), the constraints of first order primordial scalar perturbations are much weaker than the constraints on large scales \cite{Bringmann:2011ut}. The amplitude of the small-scale primordial scalar perturbations can be large enough to generate the primordial black holes, and it has close relations with the scalar induced gravitational waves \cite{Zhou:2021vcw,Ananda:2006af,Kohri:2018awv,Dalianis:2020gup,Baumann:2007zm,Inomata:2018epa,Byrnes:2018txb,Cai:2018dig,Yuan:2019udt,Inomata:2020cck,Sasaki:2018dmp,Alabidi:2012ex,Domenech:2021ztg,Balaji:2022rsy,Romero-Rodriguez:2021aws}. Similarly, the cosmological background of higher order vector modes can be generated by the first order scalar perturbations \cite{Lu:2007cj,Mollerach:2003nq,Christopherson:2009bt,Christopherson:2010ek,Lu:2008ju,Saga:2017hft}. These higher order vector modes will affect many cosmological observations, such as CMB polarization \cite{Mollerach:2003nq,Kamionkowski:1996zd}, redshift-space distortions \cite{Smith:2007sb}, and weak lensing \cite{Durrer:1994uu,Yamauchi:2012bc,Saga:2015apa}. 

In the study of second order scalar induced gravitational waves, it is convenient to consider a monochromatic primordial power spectrum $P_{\phi}=Ak_*\delta\left(k-k_*\right)$. However, it is not the case for the second order vector modes. As mentioned in Ref.~\cite{Lu:2007cj}, the second order vector modes can not be generated by a monochromatic primordial power spectrum. More precisely, if we calculate the power spectrum of second order vector modes $\mathcal{P}^{(2)}_V$ in terms of a monochromatic primordial power spectrum, we will obtain $\mathcal{P}^{(2)}_V=0$. This property of second order vector modes is very different from the second order scalar perturbations and the second order gravitational waves.  

In this paper, we investigate the second order and the third order vector modes systematically. We conclude that the vector modes can be generated by a monochromatic primordial power spectrum at third order. In this case, the first order and the second order vector modes do not exist. The third order vector mode is the first non-trivial order. The explicit expressions of two point functions $\langle V^{(n),\lambda}V^{(n),\lambda'} \rangle$$\left(n=2,3\right)$ and power spectra corresponded are presented. It shows that the third order vector modes sourced by the second order scalar perturbations dominate the two point functions $\langle V^{(3),\lambda}V^{(3),\lambda'} \rangle$ and power spectrum corresponded of third order vector modes. 

This paper is organized as follows. In Sec.~\ref{sec:2}, we study the kernel function and the power spectrum of the second order vector modes. We show that the second order vector modes can not be generated by a monochromatic primordial power spectrum. In Sec.~\ref{sec:3}, we investigate the kernel functions and the power spectrum of the third order vector modes. The explicit expressions of the power spectrum of the third order vector modes are presented. The conclusions and discussions are summarized in Sec.~\ref{sec:4}.

\section{Second order vector perturbations}\label{sec:2}
In flat Friedmann-Robertson-Walker (FRW) spacetime, the background metric is given by
\begin{equation}
	g_{\mu \nu}^{(0)} \mathrm{d} x^{\mu} \mathrm{d} x^{\nu}=a^{2}(\eta)\left(-\mathrm{d} \eta^{2}+\delta_{i j} \mathrm{d} x^{i} \mathrm{d} x^{j}\right) \ ,
\end{equation}
where $\eta$ is the conformal time. The second order metric perturbations in Newtonian gauge is
\begin{equation}
	\begin{aligned}
		\mathrm{d} s^{2}&=-a^{2}\Bigg[\left(1+2 \phi^{(1)}\right) \mathrm{d} \eta^{2}+ V_i^{(2)} \mathrm{d} \eta \mathrm{d} x^{i}+\left(1-2 \psi^{(1)}\right) \delta_{i j}\mathrm{d} x^{i} \mathrm{d} x^{j}\Bigg] \ ,
	\end{aligned}
\end{equation}
where $\phi^{(1)}$ and $\psi^{(1)}$ are the first order scalar perturbations,  $V^{(2)}_i$ is the second order vector perturbation. In this section, we consider second order vector mode $V^{(2)}_i$ generated by the first order scalar perturbations. So that we neglect the first order vector perturbation $V^{(1)}_i$, the second order scalar perturbations $\phi^{(2)}$ and $\psi^{(2)}$, and the tensor perturbations $h^{(n)}_{ij}$.

\subsection{Second order kernel function}\label{sec:2.1}
\begin{figure}
    \centering
    \includegraphics[scale = 0.8]{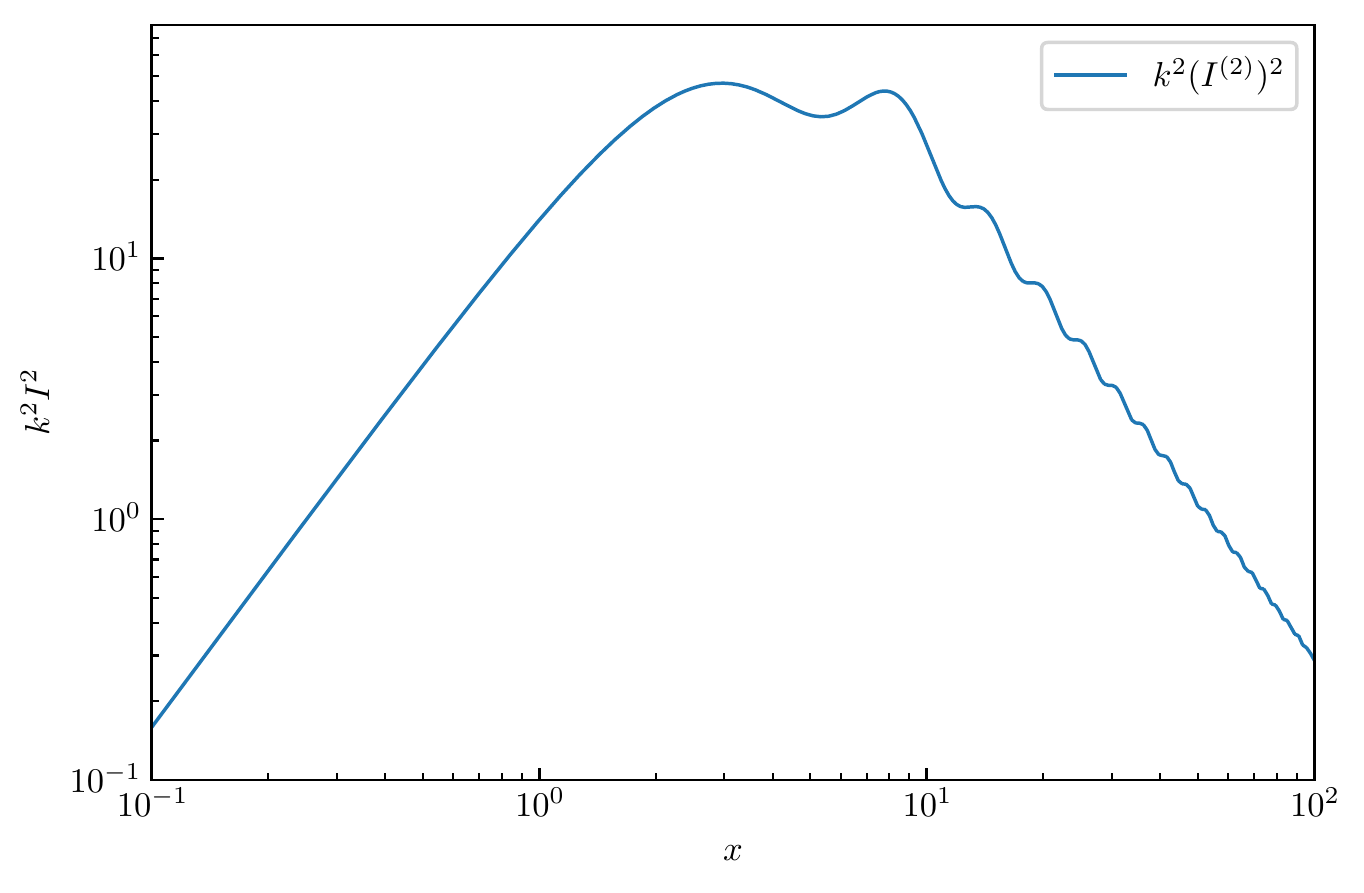}
    \caption{The square of the second order vector kernel function. Here we have set $u=v=1$.}\label{fig:I2}
\end{figure}

We consider second order vector mode $V^{(2)}_i$ generated by the first order scalar perturbations. The equation of motion of second order vector perturbation is 
\begin{equation}\label{eq:eqV2}
	\begin{aligned}
		V_{l}^{(2)'}(\eta,\mathbf{x})+2 \mathcal{H} V_{l}^{(2)}(\eta,\mathbf{x}) =-4 \Delta^{-1} \mathcal{T}_{l}^{r} \partial^{s} \mathcal{S}^{(2)}_{rs}(\eta,\mathbf{x}) \ ,
	\end{aligned}
\end{equation}
where the source term is given by
\begin{equation}
	\begin{aligned}
		\Delta^{-1} \mathcal{T}_{l}^{r} \partial^{s} \mathcal{S}^{(2)}_{rs}(\eta,\mathbf{x})=& \Delta^{-1} \mathcal{T}_{l}^{r} \partial^{s}\Bigg(\partial_{r} \phi^{(1)} \partial_{s} \phi^{(1)}-\frac{1}{ \mathcal{H}}\left(\partial_{r} \phi^{(1)'} \partial_{s} \phi^{(1)}+\partial_{r} \phi^{(1)}  \partial_{s} \phi^{(1)'}\right)\\
		&+4 \phi^{(1)} \partial_{r} \partial_{s} \phi^{(1)}-\frac{1}{ \mathcal{H}^{2}} \partial_{r} \phi^{(1)'} \partial_{s}  \phi^{(1)'}\Bigg) \ .
	\end{aligned}
\end{equation}
The details of Eq.~(\ref{eq:eqV2}) are given in Appendix~\ref{sec:AB1}.  After making use of the Fourier transformation, we obtain the equation of motion of second order vector perturbation in momentum space
\begin{equation}\label{eq:eV2}
	\begin{aligned}
		V^{\lambda,(2)'}(\eta,\mathbf{k})+2 \mathcal{H} V^{\lambda,(2)}(\eta,\mathbf{k}) =4 \mathcal{S}_V^{\lambda,(2)}(\eta,\mathbf{k}) \ ,
	\end{aligned}
\end{equation}
where $V^{\lambda,(2)}(\eta,\mathbf{k})=e^{\lambda,l}(\mathbf{k})V_{l}^{(2)}(\eta,\mathbf{k})$, $e^{\lambda,l}(\mathbf{k})$ is the polarization vectors with respect to $\mathbf{k}$, which satisfies $\sum_{\lambda}	e^{\lambda,m}(\mathbf{k})e^{\lambda,r}(\mathbf{k})+\frac{k^mk^r}{k^2}=\delta^{mr}$, $e^{\lambda,l}(\mathbf{k})e^{\lambda'}_l(\mathbf{k})=\delta^{\lambda\lambda'}$, and $e^{\lambda,l}(\mathbf{k})k_l=0$. The source term can be written as
\begin{equation}\label{eq:SV}
	\begin{aligned}
		\mathcal{S}_V^{\lambda,(2)}(\eta,\mathbf{k})&=i\frac{p^se^{\lambda,r}(\mathbf{k})}{k^2} \mathcal{S}_{rs}(\eta,\mathbf{k})\\
		&=\int\frac{d^3p}{(2\pi)^{3/2}}i\frac{k^se^{\lambda,r}(\mathbf{k})}{k^2}p_rp_s\Bigg(  3\phi^{(1)}(\mathbf{k}-\mathbf{p})  \phi^{(1)}(\mathbf{p})+\frac{2}{ \mathcal{H}} \phi^{(1)'}(\mathbf{k}-\mathbf{p})  \phi^{(1)}(\mathbf{p})\\
		&+\frac{1}{ \mathcal{H}^{2}}\phi^{(1)'}(\mathbf{k}-\mathbf{p})   \phi^{(1)'}(\mathbf{p})
		\Bigg) \ .
	\end{aligned}
\end{equation}
We rewrite the first order scalar perturbations in the form of
\begin{equation}
	\psi(\eta,\mathbf{p}) = \phi(\eta,\mathbf{p}) = \Phi_{\mathbf{p}} T_\phi(p \eta)~,
\end{equation}
where $\Phi_{\mathbf{p}}$ is initial value originated from primordial curvature perturbation, and the transfer function $T_{\phi}(y)=\frac{9}{y^{2}}\left(\frac{\sqrt{3}}{y} \sin \left(\frac{y}{\sqrt{3}}\right)-\cos \left(\frac{y}{\sqrt{3}}\right)\right) $.
Therefore, the formal expression of the second order vector mode can be written as
\begin{equation}\label{eq:V2}
	\begin{aligned}
		V^{\lambda,(2)}(\eta,\mathbf{k})=\int\frac{d^3p}{(2\pi)^{3/2}}i\frac{k^se^{\lambda,r}(\mathbf{k})}{k^2}p_rp_s I^{(2)}_V\left(|k-p|,p,\eta  \right)\Phi_{\mathbf{k}-\mathbf{p}}\Phi_{\mathbf{p}} \ .
	\end{aligned}
\end{equation}
We define $k \equiv |\mathbf{k}|$, $p\equiv |\mathbf{p}|=vk$, $|\mathbf{k}-\mathbf{p}|=uk$, and $x=k\eta$. Then, the corresponding second order kernel function is 
\begin{equation}\label{eq:IV}
	\begin{aligned}
		I_{V}^{(2)}(u,v,x)=\frac{4}{k}\int_0^{x} f_V^{(2)}(u,v,\bar{x})\frac{\bar{x}^2}{x^2} d\bar{x} \ ,
	\end{aligned}
\end{equation}
where
\begin{equation}\label{eq:fV2}
	\begin{aligned}
		f_V(u,v,x)= 3T_{\phi}(ux)  T_{\phi}(vx)+2ux \frac{d}{d(ux)}T_{\phi}(ux)  T_{\phi}(vx) +uvx^2\frac{d}{d(ux)}T_{\phi}(ux)  \frac{d}{d(vx)}T_{\phi}(vx) \ .
	\end{aligned}
\end{equation}
We present the second order kernel function $I_{V}^{(2)}(u=1,v=1,x)$ as a function of $x=k\eta$ in Fig.~\ref{fig:I2}.

\subsection{Second order power spectrum}\label{sec:2.2}
The power spectrum of second order vector mode $\mathcal{P}_V(\eta,k)$ is defined as 
\begin{equation}\label{eq:PV2}
	\begin{aligned}
		&\delta_{\lambda\lambda'}\langle V^{\lambda,(2)}(\eta,\mathbf{k})V^{\lambda',(2)}(\eta,\mathbf{k}')\rangle=2\delta(\mathbf{k}+\mathbf{k}')\frac{2\pi^2}{k^3}\mathcal{P}^{(2)}_{V}(\eta, k) \ .
	\end{aligned}
\end{equation}
Substituting Eq.~(\ref{eq:V2}) into Eq.~(\ref{eq:PV2}), we obtain
\begin{equation}\label{eq:Power2}
	\begin{aligned}
		&2\delta(\mathbf{k}+\mathbf{k}')\frac{2\pi^2}{k^3}\mathcal{P}^{(2)}_{V}(\eta, k)
		=-\delta_{\lambda\lambda'}\int\frac{d^3pd^3p'}{(2\pi)^3}\frac{k^se^{\lambda,r}(\mathbf{k})}{k^2}\frac{k'^le^{\lambda',m}(\mathbf{k}')}{k'^2}p_rp_sp'_lp'_m \\
		&\times I^{(2)}_V\left(|k-p|,p,\eta  \right)I^{(2)}_V\left(|k'-p'|,p',\eta  \right) \langle \Phi_{\mathbf{k}-\mathbf{p}}\Phi_{\mathbf{p}} \Phi_{\mathbf{k}'-\mathbf{p}'}\Phi_{\mathbf{p}'} \rangle  \ ,
	\end{aligned}
\end{equation}
where the minus sign comes from the $(i)^2$. The integral and measure $\int\frac{d^3pd^3p'}{(2\pi)^3}$ can be simplified as the integrals of $u$ and $v$ with the explicit upper and lower limits in terms of a given coordinate system. And it is convenient to set $\mathbf{k}/|\mathbf{k}|=\left(0,0,1\right)$ in the calculations of integral and measure. The polynomial of momentums $\frac{k^se^{\lambda,r}(\mathbf{k})}{k^2}\frac{k'^le^{\lambda',m}(\mathbf{k}')}{k'^2}p_rp_sp'_lp'_m$ comes from the decomposition operator of vector mode $\Delta^{-1} \mathcal{T}_{l}^{r} \partial^{s}$ in momentum space and the polarization vectors $e^{\lambda,l}(\mathbf{k})$. It can be simplified as a polynomial of $u$, $v$, and $k$.
The product of second order kernel functions $I^{(2)}(|\mathbf{k}-\mathbf{p}|, p,\eta)I^{(2)}(|\mathbf{k}'-\mathbf{p}'|,p',\eta)$ have been studied in Sec.~\ref{sec:2.1}. The four point function $\langle \Phi_{\mathbf{k}-\mathbf{p}}\Phi_{\mathbf{p}} \Phi_{\mathbf{k}'-\mathbf{p}'}\Phi_{\mathbf{p}'} \rangle$ can be simplified in terms of Wick's theorem in Appendix~\ref{sec:C4}. It is noticed that the left hand side of Eq.~(\ref{eq:Power2}) and the last line of Eq.~(\ref{eq:Wick2}) both contain the three dimensional delta function $\delta\left(\mathbf{k}+\mathbf{k}' \right)$. We can integral over $\mathbf{k}'$ to obtain the expression of $\mathcal{P}_V^{(2)}\left(\eta,k \right)$. The Wick's expansion in Eq.~(\ref{eq:Wick2}) also contains the summation of two delta functions $\left(\delta\left(\mathbf{p}+\mathbf{p}^{\prime}\right)+\delta\left(\mathbf{k}-\mathbf{p}+\mathbf{p}^{\prime}\right)\right)$. We can integral over $\mathbf{p}'$ to obtain the substitutions $\mathbf{p}' \to -\mathbf{p}$ and $\mathbf{p}' \to \mathbf{p}-\mathbf{k}$. The explicit expression of the second order power spectrum $\mathcal{P}_V^{(2)}\left(\eta, k\right)$ is 
\begin{equation}\label{eq:PowV2}
	\begin{aligned}
		\mathcal{P}_V^{(2)}=\frac{1}{4} \int_{0}^{\infty} \mathrm{d} v \int_{|1-v|}^{1+v} \mathrm{~d} u   I^{(2)}_V\left(u,v,x  \right)\sum_{a=1}^{2}\mathbb{P}_a\left( u,v\right) I^{(2)}_{V,a}\left(u,v,x  \right)k^2  \mathcal{P}_{\Phi}(k u) \mathcal{P}_{\Phi}(k v) \ ,
	\end{aligned}
\end{equation}
where 
\begin{equation}
	\begin{aligned}
		 &\mathbb{P}_1\left( u,v \right)=\frac{\left( 1+v^2-u^2 \right)\left(4 v^{2}-\left(1+v^{2}-u^{2}\right)^{2}\right)}{16u^2v^2} \ , \\ 
		 &\mathbb{P}_2\left( u,v \right)=\frac{\left(\left( 1+v^2-u^2 \right)^2-2\left(1+v^2-u^2\right)\right)\left(4 v^{2}-\left(1+v^{2}-u^{2}\right)^{2}\right)}{16v^2u^2} \ .
	\end{aligned}
\end{equation}
The summation of index $a$ in Eq.~(\ref{eq:PowV2}) comes from the two terms in Wick's expansions of four point function. The calculation details are given in Appendix~\ref{sec:D1}. One can use Eq.~(\ref{eq:PowV2}) to calculate the power spectrum of second order vector mode for a given primordial power spectrum $\mathcal{P}_{\Phi}(k)$. 

Here, we consider a monochromatic primordial power spectrum $\mathcal{P}_{\Phi}(k)=Ak_*\delta\left(k-k_*\right)$. Then Eq.~(\ref{eq:PowV2}) can be simplified as
\begin{equation}\label{eq:PowV21}
	\begin{aligned}
			\mathcal{P}_V^{(2)}=\frac{A^2}{2\tilde{k}^2} \int_{0}^{\infty} \mathrm{d} v \int_{|1-v|}^{1+v} \mathrm{~d} u  \left(kI^{(2)}_V\left(u,v,x  \right)\right)^2\left(\delta\left(u-\frac{1}{\tilde{k}}\right)\delta\left(v-\frac{1}{\tilde{k}}\right) \right) \sum_{a=1}^{2}\mathbb{P}_a\left( u,v\right)  \ .
	\end{aligned}
\end{equation}
Integrating over $u$ and $v$, we obtain the substitutions $u=v=\frac{k_*}{k}=\frac{1}{\tilde{k}}$. It is not difficult to find that the summation of the polynomials is equal to zero, namely $\sum_{a=1}^2\mathbb{P}_a\left( \frac{1}{\tilde{k}},\frac{1}{\tilde{k}}\right)=0$. So that we conclude that the second order vector mode $V^{\lambda,(2)}$ can not be generated by a monochromatic primordial power spectrum \cite{Lu:2007cj}.

\section{Third order vector perturbations}\label{sec:3}
In this section, we consider the third order vector perturbations. The third order metric perturbations in Newtonian gauge take the form of
\begin{equation}
	\begin{aligned}
		\mathrm{d} s^{2}&=-a^{2}\Bigg[\left(1+2 \phi^{(1)}+ \phi^{(2)}\right) \mathrm{d} \eta^{2}+\left( V_i^{(2)}+\frac{1}{3} V_{i}^{(3)} \right) \mathrm{d} \eta \mathrm{d} x^{i} \\
		&+\left(\left(1-2 \psi^{(1)}- \psi^{(2)}\right) \delta_{i j}+\frac{1}{2} h_{i j}^{(2)}\right)\mathrm{d} x^{i} \mathrm{d} x^{j}\Bigg] \ ,
	\end{aligned}
\end{equation}
where $\phi^{(n)}$ and $\psi^{(n)}$$(n=1,2)$ are the $n$-order scalar perturbations,  $V^{(n)}_i$$(n=2,3)$ are the $n$-order vector perturbation, and $h_{i j}^{(2)}$ is the second order tensor perturbations. 

\subsection{Third order kernel function}\label{sec:3.1}
We study the equation of motion and the kernel functions of the third order vector mode $V^{(3)}_i$ in this section. The equation of motion of the third order vector mode is rewritten as follows,
\begin{equation}\label{eq:V3}
		V_{i}^{(3)'}(\eta,\mathbf{x})+2 \mathcal{H} V_{i}^{(3)}(\eta,\mathbf{x}) =-12 \Delta^{-1} \mathcal{T}_{i}^{l} \partial^{m} \mathcal{S}^{(3)}_{l m}(\eta,\mathbf{x}) \ ,
\end{equation}
where the expression of traceless operator is $\mathcal{T}_{i}^{l}=\delta_{i}^{l}-\partial^{l} \Delta^{-1} \partial_{i}$. The  details of the decomposed operators are shown in Appendix~\ref{sec:AA}. The explicit details of Eq.~(\ref{eq:V3}) are given in Appendix~\ref{sec:AB2}.
The source term $S_{lm}^{(3)}(\eta,\mathbf{x})$ can be divided into four parts,
\begin{equation}
	S_{lm}^{(3)}(\eta,\mathbf{x})=S_{lm,1}^{(3)}(\eta,\mathbf{x})+S_{lm,2}^{(3)}(\eta,\mathbf{x})+S_{lm,3}^{(3)}(\eta,\mathbf{x})+S_{lm,4}^{(3)}(\eta,\mathbf{x}) \ ,
\end{equation}
where $S_{lm,1}^{(3)}(\eta,\mathbf{x})$ is composed of the first order scalar perturbation $\phi^{(1)}$,
\begin{equation}
	\begin{aligned}
		S_{lm,1}^{(3)}(\eta,\mathbf{x})&=12\phi^{(1)}\partial_l\phi^{(1)}\partial_m\phi^{(1)}-\frac{4}{\mathcal{H}}\phi^{(1)'}\partial_l\phi^{(1)}\partial_m\phi^{(1)}+\frac{2}{3\mathcal{H}^2}\Delta\phi^{(1)}\partial_l\phi^{(1)}\partial_m\phi^{(1)}\\
		&+\frac{2}{3\mathcal{H}^4}\Delta\phi^{(1)}\partial_l\phi^{(1)'}\partial_m\phi^{(1)'}-\frac{3}{\mathcal{H}^2}\phi^{(1)'}\partial_l\phi^{(1)'}\partial_m\phi^{(1)}-\frac{3}{\mathcal{H}^2}\phi^{(1)'}\partial_m\phi^{(1)'}\partial_l\phi^{(1)}\\
		&+\frac{2}{3\mathcal{H}^3}\Delta\phi^{(1)}\partial_l\phi^{(1)'}\partial_m\phi^{(1)}+\frac{2}{3\mathcal{H}^3}\Delta\phi^{(1)}\partial_m\phi^{(1)'}\partial_l\phi^{(1)}\\
		&-\frac{2}{\mathcal{H}^3}\phi^{(1)'}\partial_l\phi^{(1)'}\partial_m\phi^{(1)'}-\frac{4}{\mathcal{H}^2}\phi^{(1)}\partial_l\phi^{(1)'}\partial_m\phi^{(1)'} \ .
	\end{aligned}\label{Z5}
\end{equation}
The source term $S_{lm,2}^{(3)}(\eta,\mathbf{x})$ is composed of the first order scalar perturbation $\phi^{(1)}$ and the second order tensor perturbation $h_{lm}^{(2)}$,
\begin{equation}
	\begin{aligned}
		S_{lm,2}^{(3)}(\eta,\mathbf{x})&=-\frac{1}{2}\phi^{(1)}\left( h_{lm}^{(2)''}+2 \mathcal{H}  h_{lm}^{(2)'}-\Delta h_{lm}^{(2)}\right)-\phi^{(1)}\Delta h_{lm}^{(2)}-\phi^{(1)'}\mathcal{H}h_{lm}^{(2)}-\frac{1}{3}\Delta \phi^{(1)}h_{lm}^{(2)}\\
		&-\partial^b \phi^{(1)}\partial_b h_{lm}^{(2)} \ .
	\end{aligned} 
\end{equation}
The source term $S_{lm,3}^{(3)}(\eta,\mathbf{x})$ is composed of the first order scalar perturbation $\phi^{(1)}$ and the second order vector perturbation $V_l^{(2)}$, 
\begin{equation}
	\begin{aligned}
		S_{lm,3}^{(3)}(\eta,\mathbf{x})&=\phi^{(1)}\partial_l\left(V_m^{(2)'}+2 \mathcal{H}V_m^{(2)} \right)+\phi^{(1)}\partial_m\left(V_l^{(2)'}+2 \mathcal{H}V_l^{(2)} \right)+\phi^{(1)'}\left(\partial_lV_m^{(2)}+\partial_mV_l^{(2)}\right)\\
		&-\frac{\phi^{(1)}}{8\mathcal{H}}\left(\partial_m\Delta V_l^{(2)}+\partial_l\Delta V_m^{(2)}\right)-\frac{\phi^{(1)'}}{8\mathcal{H}^2}\left(\partial_m\Delta V_l^{(2)}+\partial_l\Delta V_m^{(2)}\right) \ .
	\end{aligned}
\end{equation}
And the source term $S_{lm,4}^{(3)}(\eta,\mathbf{x})$ is composed of the first order scalar perturbation $\phi^{(1)}$ and the second order scalar perturbations $\phi^{(2)}$ and $\psi^{(2)}$, 
\begin{equation}
	\begin{aligned}
		S_{lm,4}^{(3)}(\eta,\mathbf{x})&=\frac{1}{\mathcal{H}}\left(\phi^{(1)}\partial_l\partial_m\psi^{(2)'}\right)+\frac{1}{\mathcal{H}}\left(\phi^{(1)'}\partial_l\partial_m\phi^{(2)}\right)+\frac{1}{\mathcal{H}^2}\left(\phi^{(1)'}\partial_l\partial_m\psi^{(2)'}\right)\\
		&+3\left(\phi^{(1)}\partial_l\partial_m\phi^{(2)}\right) \ .
	\end{aligned}\label{Z8}
\end{equation}
The second order perturbations $h_{lm}^{(2)}$,  $V_l^{(2)}$ and $\phi^{(2)}$ and $\psi^{(2)}$ are induced by the first order scalar perturbations $\psi^{(1)}(=\phi^{(1)})$. At the next iteration, the first order scalar, the second order scalar, vector and tensor perturbations all induce the third order vector perturbation. The kernel functions of the second order perturbations have been studied in previous work \cite{Zhou:2021vcw}. It is similar to the calculations of the second order vector mode. In order to solve the equations of motion of third order vector mode, we rewrite Eq.~(\ref{eq:V3})  in momentum space as
\begin{equation}\label{eq:V31}
	\begin{aligned}
		V^{\lambda,(3)'}(\eta,\mathbf{p})+2 \mathcal{H} V^{\lambda,(3)}(\eta,\mathbf{p}) =12 \sum^{4}_{j=1} \mathcal{S}_j^{\lambda,(3)}(\eta,\mathbf{k}) ~,
	\end{aligned}
\end{equation}
where $V^{\lambda,(3)}(\eta,\mathbf{k})=e^{\lambda,l}(\mathbf{k})V_{l}^{(3)}(\eta,\mathbf{k})$, $e^{\lambda,l}(\mathbf{k})$ is the polarization vectors with respect to $\mathbf{k}$, which satisfies $\sum_{\lambda}	e^{\lambda,m}(\mathbf{k})e^{\lambda,r}(\mathbf{k})+\frac{k^mk^r}{k^2}=\delta^{mr}$, $e^{\lambda,l}(\mathbf{k})e^{\lambda'}_l(\mathbf{k})=\delta^{\lambda\lambda'}$, and $e^{\lambda,l}(\mathbf{k})k_l=0$.
Unlike the second order tensor perturbation, the third order vector mode have four types of source terms $S_i^{\lambda,(3)}(\eta,\mathbf{k}),(i=1,2,3,4)$ (Eqs.~(\ref{Z5})--(\ref{Z8})). We rewrite the source terms $S_i^{\lambda,(3)}(\eta,\mathbf{k})$ in terms of initial value of the first order scalar perturbations $\Phi_{\mathbf{p}}$ in momentum space, i.e.,
\begin{eqnarray}
	S_1^{\lambda,(3)}(\eta,\mathbf{k})&=&\int\frac{d^3p}{(2\pi)^{3/2}}\int\frac{d^3q}{(2\pi)^{3/2}}i\frac{k^le^{\lambda,m}(\mathbf{k})}{k^2}(p_l-q_l)q_m\Phi_{\mathbf{k}-\mathbf{p}} \Phi_{\mathbf{p}-\mathbf{q}} \Phi_{\mathbf{q}}\nonumber\\
	& &\times
	f_1^{(3)}(|\mathbf{k}-\mathbf{p}|,|\mathbf{p}-\mathbf{q}|,\mathbf{q},\eta) \ , \label{eq:S31}\\
	S_2^{\lambda,(3)}(\eta,\mathbf{k})&=&\int\frac{d^3p}{(2\pi)^{3/2}}\int\frac{d^3q}{(2\pi)^{3/2}}i\frac{k^le^{\lambda,m}(\mathbf{k})}{k^2}\Lambda^{r s}_{l m}(\textbf{p})q_rq_s\Phi_{\mathbf{k}-\mathbf{p}} \Phi_{\mathbf{p}-\mathbf{q}} \Phi_{\mathbf{q}}\nonumber\\
	& &\times
	f^{(3)}_2(|\mathbf{k}-\mathbf{p}|,|\mathbf{p}-\mathbf{q}|,\mathbf{q},\eta) \ ,  \label{eq:S32} \\
	S_3^{\lambda,(3)}(\eta,\mathbf{k})&=&\int\frac{d^3p}{(2\pi)^{3/2}}\int\frac{d^3q}{(2\pi)^{3/2}}i\frac{k^le^{\lambda,m}(\mathbf{k})}{k^2}\left(\mathcal{T}^r_m(\textbf{p}) p_l+\mathcal{T}^r_l(\textbf{p})p_m\right)\frac{p^s}{p^2}q_rq_s\Phi_{\mathbf{k}-\mathbf{p}} \Phi_{\mathbf{p}-\mathbf{q}} \Phi_{\mathbf{q}} \nonumber\\
	& &\times f^{(3)}_3(|\mathbf{k}-\mathbf{p}|,|\mathbf{p}-\mathbf{q}|,\mathbf{q},\eta) \ ,\label{eq:S33} \\
	S_4^{\lambda,(3)}(\eta,\mathbf{k})&=&\int\frac{d^3p}{(2\pi)^{3/2}}\int\frac{d^3q}{(2\pi)^{3/2}}i\frac{k^le^{\lambda,m}(\mathbf{k})}{k^2}p_lp_m\Phi_{\mathbf{k}-\mathbf{p}} \Phi_{\mathbf{p}-\mathbf{q}} \Phi_{\mathbf{q}}\nonumber\\
	& &\times
	f^{(3)}_4(|\mathbf{k}-\mathbf{p}|,|\mathbf{p}-\mathbf{q}|,\mathbf{q},\eta) \ , \label{eq:S34}
\end{eqnarray}
and the expression of $f_i^{(3)}(|\mathbf{k}-\mathbf{p}|,|\mathbf{p}-\mathbf{q}|,\mathbf{q},\eta)$ are given by,
\begin{equation}
	\begin{aligned}
		f_1^{(3)}(u,\bar{u},\bar{v},x,y)&=12T_{\phi}(ux) T_{\phi}(\bar{u}y) T_{\phi}(\bar{v}y)-4ux\frac{d}{d(ux)}T_{\phi}(ux) T_{\phi}(\bar{u}y) T_{\phi}(\bar{v}y)\\
		&-\frac{2u^2x^2}{3}T_{\phi}(ux) T_{\phi}(\bar{u}y) T_{\phi}(\bar{v}y)-6\bar{u}uxy\frac{d}{d(ux)}T_{\phi}(ux) \frac{d}{d(\bar{u}y)}T_{\phi}(\bar{u}y) T_{\phi}(\bar{v}y)\\
		&-\frac{4u^2\bar{u}x^2y}{3}T_{\phi}(ux)\frac{d}{d(\bar{u}y)} T_{\phi}(\bar{u}y) T_{\phi}(\bar{v}y)-4\bar{u}\bar{v}y^2T_{\phi}(ux) \frac{d}{d(\bar{u}y)}T_{\phi}(\bar{u}y) \frac{d}{d(\bar{v}y)}T_{\phi}(\bar{v}y)\\
		&-2\bar{u}\bar{v}uy^2x\frac{d}{d(ux)}T_{\phi}(ux) \frac{d}{d(\bar{u}y)}T_{\phi}(\bar{u}y) \frac{d}{d(\bar{v}y)}T_{\phi}(\bar{v}y)\\
		&-\frac{2u^2\bar{u}\bar{v}x^2y^2}{3}T_{\phi}(ux) \frac{d}{d(\bar{u}y)}T_{\phi}(\bar{u}y) \frac{d}{d(\bar{v}y)}T_{\phi}(\bar{v}y)\ ,
	\end{aligned}
\end{equation}

\begin{equation}
	\begin{aligned}
		f_2^{(3)}(u,\bar{u},\bar{v},x,y)=&-6T_{\phi}(ux) T_{\phi}(\bar{u}y) T_{\phi}(\bar{v}y)-4\bar{u}yT_{\phi}(ux) \frac{d}{d(\bar{u}y)}T_{\phi}(\bar{u}y) T_{\phi}(\bar{v}y)\\
		&-T_{\phi}(ux)p^2I_{h}^{(2)}(\bar{u},\bar{v},y)-2\bar{u}\bar{v}y^2T_{\phi}(ux) \frac{d}{d(\bar{u}y)}T_{\phi}(\bar{u}y) \frac{d}{d(\bar{v}y)}T_{\phi}(\bar{v}y)\\
		&+\frac{u}{v^2x}\frac{d}{d(ux)}T_{\phi}(ux)p^2I_{h}^{(2)}(\bar{u},\bar{v},y)-\frac{u^2}{3v^2} T_{\phi}(ux)p^2I_{h}^{(2)}(\bar{u},\bar{v},y)\\
		&-\frac{1-u^2-v^2}{2v^2} T_{\phi}(ux)p^2 I_{h}^{(2)}(\bar{u},\bar{v},y) \ ,
	\end{aligned}\label{Z15}
\end{equation}

\begin{equation}
	\begin{aligned}
		f_3^{(3)}(u,\bar{u},\bar{v},x,y)&=\frac{u}{v}\frac{d}{d(ux)}T_{\phi}(ux)pI_V^{(2)}(\bar{u},\bar{v},y)+\frac{y}{8}T_{\phi}(ux)pI_V^{(2)}(\bar{u},\bar{v},y)\\
		&+\frac{xuy}{8}\frac{d}{d(ux)}T_{\phi}(ux)pI_V^{(2)}(\bar{u},\bar{v},y)-4T_{\phi}(ux) T_{\phi}(\bar{u}y) T_{\phi}(\bar{v}y)\\
		&+8\bar{u}y T_{\phi}(ux) \frac{d}{d(\bar{u}y)}T_{\phi}(\bar{u}y) T_{\phi}(\bar{v}y)+16T_{\phi}(ux) T_{\phi}(\bar{u}y) T_{\phi}(\bar{v}y)\\
		&+4\bar{u}\bar{v}y^2T_{\phi}(ux) \frac{d}{d(\bar{u}y)}T_{\phi}(\bar{u}y) \frac{d}{d(\bar{v}y)}T_{\phi}(\bar{v}y) \ ,
	\end{aligned}
\end{equation}

\begin{equation}
	\begin{aligned}
		f_4^{(3)}(u,\bar{u},\bar{v},\eta)=&y\left(T_{\phi}(ux)\frac{\partial}{\partial y}I^{(2)}_{\psi}(\bar{u},\bar{v},y)\right)+uxy\left(\frac{d}{d(ux)}T_{\phi}(ux)\frac{\partial}{\partial y}I^{(2)}_{\psi}(\bar{u},\bar{v},y)\right)\\
		&+ux\left(\frac{d}{d(ux)}T_{\phi}(ux)(I^{(2)}_{\psi}(\bar{u},\bar{v},y)+f^{(2)}_{\phi}(\bar{u},\bar{v},y))\right)\\
		&+3\left(T_{\phi}(ux)(I^{(2)}_{\psi}(\bar{u},\bar{v},y)+f^{(2)}_{\phi}(\bar{u},\bar{v},y))\right) \ ,
	\end{aligned} \label{Z18}
\end{equation}
where we have set $|\mathbf{k}-\mathbf{p}|=uk$, $p=vk$, $|\mathbf{k}-\mathbf{q}|=wk$, $|\mathbf{p}-\mathbf{q}|=\bar{u}p$, $q=\bar{v}p$, $x=k\eta$, and $y=p\eta$. As shown in Eqs.~(\ref{Z15})--(\ref{Z18}), we need to calculate the kernel function of the second order scalar, vector, tensor perturbations induced by the first order scalar perturbations. 

\begin{figure}
    \centering
    \includegraphics[scale = 0.8]{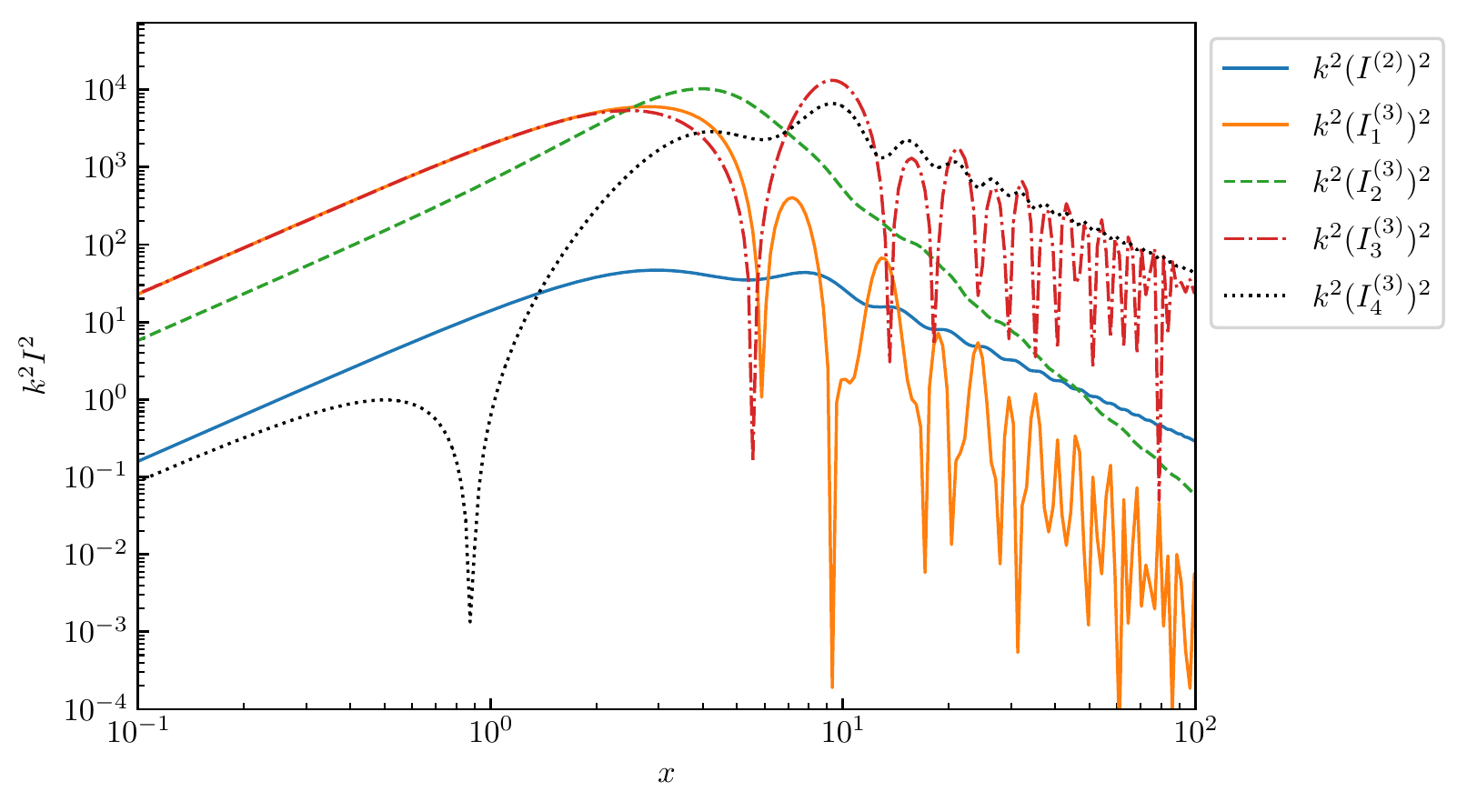}
    \caption{The squares of the kernel functions of the second order and four third order vector perturbations. Here we have set $u=v=\bar{u}=\bar{v}=1$.}\label{fig:I2_and_I3}
\end{figure}

Corresponding to four kinds of the source terms in Eqs.~(\ref{eq:S31})--(\ref{eq:S34}), we obtain the formal expression of third order vector mode 
\begin{equation}\label{eq:h30}
	\begin{aligned}
		V^{\lambda,(3)}(\eta,\mathbf{k})&=&V_1^{\lambda,(3)}(\eta,\mathbf{k})+V_2^{\lambda,(3)}(\eta,\mathbf{k})+V_3^{\lambda,(3)}(\eta,\mathbf{k})+V_4^{\lambda,(3)}(\eta,\mathbf{k}) \ ,
	\end{aligned}
\end{equation}
where $V_i^{\lambda,(3)},(i=1,2,3,4)$ are defined as
\begin{eqnarray}
	V_1^{\lambda,(3)}(\eta,\mathbf{k})&=&\int\frac{d^3p}{(2\pi)^{3/2}}\int\frac{d^3q}{(2\pi)^{3/2}}i\frac{k^le^{\lambda,m}(\mathbf{k})}{k^2}(p_l-q_l)q_m\Phi_{\mathbf{k}-\mathbf{p}} \Phi_{\mathbf{p}-\mathbf{q}} \Phi_{\mathbf{q}} \nonumber\\
	& &\times
	I_1^{(3)}(|\mathbf{k}-\mathbf{p}|,|\mathbf{p}-\mathbf{q}|,\mathbf{q},\eta) \ ,\label{eq:h31}\\
	V_2^{\lambda,(3)}(\eta,\mathbf{k})&=&\int\frac{d^3p}{(2\pi)^{3/2}}\int\frac{d^3q}{(2\pi)^{3/2}}i\frac{k^le^{\lambda,m}(\mathbf{k})}{k^2}\Lambda_{lm}^{ rs}(\mathbf{p})q_rq_s\Phi_{\mathbf{k}-\mathbf{p}} \Phi_{\mathbf{p}-\mathbf{q}} \Phi_{\mathbf{q}} \nonumber\\
	& &\times
	I^{(3)}_2(|\mathbf{k}-\mathbf{p}|,|\mathbf{p}-\mathbf{q}|,\mathbf{q},\eta) \ ,\label{eq:h32}\\
	V_3^{\lambda,(3)}(\eta,\mathbf{k})&=&\int\frac{d^3p}{(2\pi)^{3/2}}\int\frac{d^3q}{(2\pi)^{3/2}}i\frac{k^le^{\lambda,m}(\mathbf{k})}{k^2}\left(\mathcal{T}^s_m(\textbf{p}) p_l+\mathcal{T}^s_l(\textbf{p})p_m\right)\frac{p^s}{p^2}q_rq_s\Phi_{\mathbf{k}-\mathbf{p}} \Phi_{\mathbf{p}-\mathbf{q}} \Phi_{\mathbf{q}} \nonumber\\
	& &\times I^{(3)}_3(|\mathbf{k}-\mathbf{p}|,|\mathbf{p}-\mathbf{q}|,\mathbf{q},\eta) \ ,\label{eq:h33}\\
	V_4^{\lambda,(3)}(\eta,\mathbf{k})&=&\int\frac{d^3p}{(2\pi)^{3/2}}\int\frac{d^3q}{(2\pi)^{3/2}}i\frac{k^le^{\lambda,m}(\mathbf{k})}{k^2}p_lp_m\Phi_{\mathbf{k}-\mathbf{p}} \Phi_{\mathbf{p}-\mathbf{q}} \Phi_{\mathbf{q}} \nonumber\\
	& &\times
	I^{(3)}_4(|\mathbf{k}-\mathbf{p}|,|\mathbf{p}-\mathbf{q}|,\mathbf{q},\eta) \ .\label{eq:h34}
\end{eqnarray}
 Substituting Eqs.~(\ref{eq:S31})--(\ref{eq:S34}) and (\ref{eq:h31})--(\ref{eq:h34}) into Eq.~(\ref{eq:V31}), we obtain the equations of motion of kernel functions $I_i^{(3)}(u,\bar{u},\bar{v},x)$ in the form of 
\begin{equation}
	\begin{aligned}
		I_i^{(3)'}(u,\bar{u},\bar{v},x)+2 \mathcal{H}I_i^{(3)}(u,\bar{u},\bar{v},x)=12f_i^{(3)}(u,\bar{u},\bar{v},x) \ , \ (i=1,2,3,4) \ .
	\end{aligned}
\end{equation}
The kernel functions can be expressed as
\begin{equation}
	\begin{aligned}
		I_{i}^{(3)}(u,\bar{u},\bar{v},x)=\frac{12}{k} \int_{0}^{x} \mathrm{d}\bar{x} \left(\frac{\bar{x}^2}{x^2}  f_{i}^{(3)}(u,\bar{u},\bar{v},\bar{x})\right) \ , \ (i=1,2,3,4) \ .
	\end{aligned}
\end{equation}
It is similar to the second order kernel function. We present  $\left(I_i(u=1,v=1,\bar{u}=1,\bar{v}=1,x)\right)^2$, $(i,j=1,2,3,4)$ as function of $x=k\eta$ in Fig.~\ref{fig:I2_and_I3}. 
%It shows that the amplitude of $(I^{(3)}_4)^2$ is the largest. 
We notice that $I^{(3)}_4$ has the largest amplitudes for large $x$. 

\subsection{Third order power spectrum}\label{sec:3.2}
In this section, we will study the power spectrum of the third order vector mode. An explicit expression of the power spectrum $\mathcal{P}_{V}^{(3)}(\eta, k)$ is presented. Substituting Eq.~(\ref{eq:h30}) into the definition of the power spectrum , we obtain 
\begin{equation}\label{eq:PowV3}
	\begin{aligned}
		&\delta_{\lambda\lambda'}\langle V^{\lambda,(3)}(\eta,\mathbf{k})V^{\lambda',(3)}(\eta,\mathbf{k}')\rangle=\sum_{i,j=1}^{4} \langle V^{\lambda,(3)}_i(\eta,\mathbf{k})V^{\lambda',(3)}_j(\eta,\mathbf{k}')\rangle=2\delta(\mathbf{k}+\mathbf{k}')\frac{2\pi^2}{k^3}\sum_{i,j=1}^{4}\mathcal{P}^{ij}_{V}(\eta, k) \ .
	\end{aligned}
\end{equation}
The power spectrum of third order vector mode $\mathcal{P}_{V}^{(3)}(\eta, \mathbf{k})$ can be written as the sum of sixteen terms,
\begin{equation}\label{eq:p}
	\begin{aligned}
		\mathcal{P}_{V}^{(3)}(\eta, \mathbf{k})=\sum_{i,j=1}^{4}\mathcal{P}_{V}^{ij}(\eta, \mathbf{k}).
	\end{aligned}
\end{equation}
The summation of $i$ and $j$ is originated from the four kinds of source terms of the third order vector mode $V^{(3)}_l$. 
Based on the formal expressions of $V^\lambda_i(\eta,\mathbf{k})$ in Eqs.~(\ref{eq:h31})--(\ref{eq:h34}), we obtain the formal expression of power spectra $\mathcal{P}_{V}^{ij}(\eta, \mathbf{k})$ in the form of
\begin{equation}\label{eq:Ppij}
	\begin{aligned}
		\mathcal{P}_{V}^{ij}(\eta, k)&=\frac{k^3(2\pi^2)^2}{2}\int \frac{\mathrm{d}^{3} p \mathrm{~d}^{3} q\mathrm{d}^{3} p' \mathrm{~d}^{3} q'}{(2 \pi)^{6}}\mathbb{P}^{ij}(\mathbf{k},\mathbf{p},\mathbf{p}',\mathbf{q},\mathbf{q}') \mathcal{C}(\mathbf{k},\mathbf{k}',\mathbf{p},\mathbf{p}',\mathbf{q},\mathbf{q}')\\
		&\times I_i^{(3)}(|\mathbf{k}-\mathbf{p}|,|\mathbf{p}-\mathbf{q}|, p, q,\eta)I_j^{(3)}(|\mathbf{k}'-\mathbf{p}'|,|\mathbf{p}'-\mathbf{q}'|, p', q',\eta)  \ ,
	\end{aligned}
\end{equation}
where $\mathcal{C}(\mathbf{k},\mathbf{k}',\mathbf{p},\mathbf{p}',\mathbf{q},\mathbf{q}')$ is derived from  
the six-point correlation function, and its explicit expression is shown in Eq.~(\ref{eq:C}). The integral and measure $\int \frac{\mathrm{d}^{3} p \mathrm{~d}^{3} q\mathrm{d}^{3} p' \mathrm{~d}^{3} q'}{(2 \pi)^{6}}$ should be simplified as the integrals of $u$, $v$, $\bar{u}$, $\bar{v}$, and $w$ with the explicit upper and lower limits in terms of a given coordinate system. Here the calculations are similar to the calculations in the power spectrum of second order vector mode, and it will be more complicated. The polynomial of momentums $\mathbb{P}^{ij}(\mathbf{k},\mathbf{p},\mathbf{p}',\mathbf{q},\mathbf{q}')$ is originated from the decomposition operators, the polarization vectors, and the polarization tensors. We will simplify these polynomial of momentums as a polynomial of $u$, $v$, $\bar{u}$, $\bar{v}$, and $w$. The product of third order kernel functions $I_i^{(3)}(|\mathbf{k}-\mathbf{p}|,|\mathbf{p}-\mathbf{q}|, p, q,\eta)I_j^{(3)}(|\mathbf{k}'-\mathbf{p}'|,|\mathbf{p}'-\mathbf{q}'|, p', q',\eta)$ have been studied in Sec.~\ref{sec:3.1}. The Wick's expressions of six point function $\mathcal{C}(\mathbf{k},\mathbf{k}',\mathbf{p},\mathbf{p}',\mathbf{q},\mathbf{q}')$ are shown in Appendix~\ref{sec:C6}. There are six non-trivial terms in the Wick's expressions, they both contain the three dimensional delta functions of $\mathbf{p}'$ and $\mathbf{q}'$. For example, the first term in Eq.~(\ref{eq:C}) contains $\delta\left(\mathbf{p}+\mathbf{p}'\right)\delta\left(\mathbf{q}+\mathbf{q}'\right)$. When we integral over $\mathbf{p}'$ and $\mathbf{q}'$, we will obtain the substitutions $\mathbf{p}'\to -\mathbf{p}$ and $\mathbf{q}'\to -\mathbf{q}$. These substitutions will change the variables $\mathbf{p}'$ and $\mathbf{q}'$ in the polynomial $\mathbb{P}^{ij}(\mathbf{k},\mathbf{p},\mathbf{p}',\mathbf{q},\mathbf{q}')$ and the second kernel functions $I_j^{(3)}(|\mathbf{k}'-\mathbf{p}'|,|\mathbf{p}'-\mathbf{q}'|, p', q',\eta)$. The explicit expression of the power spectra $\mathcal{P}^{ij}_V$ in Eq.~(\ref{eq:Ppij}) can be written as, 
\begin{equation}\label{eq:Pex} 
	\begin{aligned}
		\mathcal{P}^{ij}_V& = \frac{1}{2 \pi} \int_0^{\infty} {\rm d} v \int_0^{\infty}{\rm d} \bar{v} \int^{1 + \bar{v} v}_{| 1 -  \bar{v} v |} {\rm d} w \int^{1 + v}_{| 1 - v |}{\rm d} u \int_{\bar{u}_-}^{\bar{u}_+} {\rm d} \bar{u} \Big\{ \frac{w}{u^2 \bar{u}^2 v \bar{v}^2 \sqrt{Y (1 - X^2)}}I^{(3)}_i \left( u, v, \bar{u}, \bar{v}, x \right)    \\
		&  \times\sum_{a=1}^{6} \mathbb{P}_{a}^{ij} \left( k, u,v,\bar{u},\bar{v},w,x \right) I^{(3),a}_j \left( u, v, \bar{u}, \bar{v},w, x \right) P_{\Phi} (k  u) P_{\Phi} (k \bar{u} v) P_{\Phi}(k \bar{v} v) \Big\}~,                        
	\end{aligned}
\end{equation}
where 
\begin{equation}
	\begin{aligned}
		\bar{u}_{\pm} & = \Bigg( 1 + \bar{v}^2 - \frac{(1 + v^2 - u^2) (1 + (\bar{v} v)^2 - w^2)}{2 v^2}   \\
		&  \pm 2 \bar{v} \sqrt{\left( 1 - \left( \frac{1 + v^2 - u^2}{2 v} \right)^2 \right) \left( 1 - \left( \frac{1 + (\bar{v}  v)^2 - w^2}{2 \bar{v} v} \right)^2 \right)}   \Bigg)^{\frac{1}{2}}~,
	\end{aligned}
\end{equation}  

\begin{equation}
	\begin{aligned}
		X & = \left(- 1 + u^2 + v^2 - 2 \bar{u}^2 v^2 +v^2   \bar{v}^2 + u^2 v^2 \bar{v}^2 - v^4\bar{v}^2 + w^2 - u^2 w^2 + v^2 w^2\right)\\
		&\times \left(1 - 2 u^2 + u^4 -   2 v^2 - 2 u^2 v^2 + v^4) (1 - 2 v^2 \bar{v}^2 + v^4 \bar{v}^4 - 2 w^2 - 2 v^2  \bar{v}^2 w^2 + w^4 \right)^{-\frac{1}{2}}~,
	\end{aligned}
\end{equation} 

\begin{equation}
	\begin{aligned}
		Y = (1 - 2 u^2 + u^4 - 2 v^2 - 2 u^2 v^2 + v^4) (1	- 2 v^2 \bar{v}^2 + v^4 \bar{v}^4 - 2 w^2 - 2	v^2 \bar{v}^2 w^2 + w^4)~.
	\end{aligned}
\end{equation} 
Some details of Eq.(\ref{eq:Pex}) are given in Appendix~\ref{sec:D2}. The summation of index $a$ in Eq.~(\ref{eq:Pex}) comes from the six terms in Wick's expansions. More precisely, the polynomials $\mathbb{P}^{ij}_a$ and the second kernel functions $I^{(3),a}_j\left( |\mathbf{k}'-\mathbf{p}'|,|\mathbf{p}'-\mathbf{q}'|, p', q',\eta \right)$ take the form of
\begin{equation}\label{Z37}
	\begin{aligned}
		\mathbb{P}^{i  j} \left( \textbf{k}, \textbf{p},  \textbf{p}',	\textbf{q},  \textbf{q}' \right)& \rightarrow \mathbb{P}^{ij}_a= \Bigg( \mathbb{P}^{i  j}	\left( \textbf{k}, \textbf{p}, - \textbf{p}, \textbf{q}, - \textbf{q}	\right) , \mathbb{P}^{i  j} \left( \textbf{k}, \textbf{p}, -	\textbf{p}, \textbf{q}, \textbf{q} - \textbf{p} \right) ,\mathbb{P}^{i		 j} \left( \textbf{k}, \textbf{p}, \textbf{p} - \textbf{q} -	\textbf{k}, \textbf{q}, - \textbf{q} \right)   \\
		&, \mathbb{P}^{i  j} \left( \textbf{k}, \textbf{p}, \textbf{q} -	\textbf{k}, \textbf{q}, \textbf{q} - \textbf{p} \right) , \mathbb{P}^{i		 j} \left( \textbf{k}, \textbf{p}, \textbf{p} - \textbf{k} -	\textbf{q}, \textbf{q}, \textbf{p} - \textbf{k} \right)   \\
		&, \mathbb{P}^{i		 j} \left( \textbf{k}, \textbf{p}, \textbf{q} - \textbf{k},	\textbf{q}, - \textbf{k} + \textbf{p} \right)\Bigg) ~,  
	\end{aligned}
\end{equation}
\begin{equation}\label{Ij}
	\begin{aligned}
		I_j^{(3)}(|\mathbf{k}'-\mathbf{p}'|&,|\mathbf{p}'-\mathbf{q}'|, p', q',\eta)  \rightarrow I^{(3),a}_j=\Bigg(  I_j^{(3)}(|\mathbf{k}-\mathbf{p}|,|\mathbf{p}-\mathbf{q}|,  p, q,\eta) , I_j^{(3)}(|\mathbf{k}-\mathbf{p}|,|\mathbf{p}-\mathbf{q}|,  p, |\mathbf{p}-\mathbf{q}|,\eta) \\
		&, I_j^{(3)}(|\mathbf{k}-\mathbf{p}|,|\mathbf{p}-\mathbf{q}|, |\textbf{p} - \textbf{q} -	\textbf{k}|, q,\eta) , I_j^{(3)}(|\mathbf{k}-\mathbf{p}|,|\mathbf{p}-\mathbf{q}|,  |\mathbf{q}-\mathbf{k}|, |\mathbf{p}-\mathbf{q}|,\eta) \\
		&, I_j^{(3)}(|\mathbf{k}-\mathbf{p}|,|\mathbf{p}-\mathbf{q}|, |\textbf{p} - \textbf{q} -	\textbf{k}|, |\mathbf{p}-\mathbf{k}|,\eta) \\
		&, I_j^{(3)}(|\mathbf{k}-\mathbf{p}|,|\mathbf{p}-\mathbf{q}|, |\textbf{q} - 	\textbf{k}|, |\mathbf{p}-\mathbf{k}|,\eta)\Bigg) ~. 
	\end{aligned}
\end{equation}

In order to obtain a specific result of $\mathcal{P}^{(3)}_{V}(\eta, \mathbf{k})$, we consider a monochromatic primordial power spectrum,
\begin{equation}\label{eq:Pphi}
	\begin{aligned}
		P_{\Phi}(k)=Ak_{*}\delta(k-k_{*}) \ .
	\end{aligned}
\end{equation}
In this case, the power spectrum $\mathcal{P}^{i j}_V$ reduce to 
\begin{eqnarray}
	\mathcal{P}^{i  j}_V(x,\tilde{k}) & = & \frac{\mathcal{A}^3 \tilde{k}^3}{2 \pi}	\Theta (3 - \tilde{k})\int_{\left| 1 - \frac{1}{\tilde{k}}		\right|}^{\min \left\{ \frac{2}{\tilde{k}}, 1 + \frac{1}{\tilde{k}}		\right\}}   {\rm d} v \nonumber \int_{w_-}^{w_+} {\rm d} w\left(\frac{ v w}{\sqrt{Y (1 - X^2)}}  \right. \\
	& &\left. \times \ I^{(3)}_i\left(u,v,\bar{u},\bar{v},x\right) \ \sum^{3}_{a=1} \mathbb{P}^{i  j}_{a} \  I^{(3),a}_j\left( u,v,\bar{u},\bar{v},w,x \right) \ \right)_{u=\frac{1}{\tilde{k}}, \bar{u}=\bar{v}=\frac{1}{v \tilde{k}}} ~, \label{eq:Z46}
\end{eqnarray}
where we have defined $\tilde{k}\equiv \frac{k}{k_*}$ and 
\begin{equation}
	\begin{aligned}
		w_{\pm} & = \Bigg(\frac{1}{2} + \frac{3}{2 \tilde{k}^2} - \frac{1}{2}
		v^2 \pm \frac{1}{2 v} \sqrt{\left( v^2 -
			\frac{4}{\tilde{k}^2} \right) \left( v^2 - \left( 1 -
			\frac{1}{\tilde{k}} \right)^2 \right) \left( v^2 - \left( 1 +
			\frac{1}{\tilde{k}} \right)^2 \right)}\Bigg)^{\frac{1}{2}}~.
	\end{aligned}
\end{equation}
In the case of monochromatic primordial power spectrum $	P_{\Phi}(k)=Ak_{*}\delta(k-k_{*})$, the summation of index $a$ reduce to three terms
\begin{equation}\label{Z3}
	\begin{aligned}
		\mathbb{P}^{i  j} \left( \textbf{k}, \textbf{p},  \textbf{p}',	\textbf{q},  \textbf{q}' \right) \rightarrow \mathbb{P}^{i  j}_{a}=\left(\mathbb{P}^{i  j}_{1} \ , \ \mathbb{P}^{i  j}_{2} \ , \ \mathbb{P}^{i  j}_{3} \right) ~,  
	\end{aligned}
\end{equation}
where the explicit expressions of $\mathbb{P}^{ij}_{a},(a=1,2,3)$ in Eq.~(\ref{Z3}) are shown in the Appendix~\ref{sec:E}. The corresponding kernel functions can be expressed as 
\begin{equation}
	\begin{aligned}
		I_j^{(3)}&(|\mathbf{k}'-\mathbf{p}'|,|\mathbf{p}'-\mathbf{q}'|, p', q',\eta)  \rightarrow I^{(3),a}_j\left( |\mathbf{k}'-\mathbf{p}'|,|\mathbf{p}'-\mathbf{q}'|, p', q',\eta \right) \\
		&=2\left(I_j^{(3),1}(k_*,k_*,  p, k_*,\eta),I_j^{(3),2}(k_*,k_*,  |\mathbf{k}-\mathbf{q}|, k_*,\eta),I_j^{(3),3}(k_*,k_*, |\textbf{p} - \textbf{q} -	\textbf{k}|, k_*,\eta) \right) ~.
	\end{aligned}
\end{equation}
Namely,
\begin{equation}\label{Ij2}
	\begin{aligned}
		&I^{(3),1}_j\left( u',v',\bar{u}',\bar{v}',x \right) =  I_j^{(3)}\left(\frac{1}{\tilde{k}},v,\frac{1}{v\tilde{k}},\frac{1}{v\tilde{k}},x\right)~, 
		\\
		& I^{(3),2}_j\left( u',v',\bar{u}',\bar{v}',x \right) = I_j^{(3)}\left(\frac{1}{\tilde{k}},w,\frac{1}{w\tilde{k}},\frac{1}{w\tilde{k}},x\right)~,  \\
		&I^{(3),3}_j\left( u',v',\bar{u}',\bar{v}',x \right)=  I_j^{(3)}\left(\frac{1}{\tilde{k}}, \left(1-v^2-w^2+3/\tilde{k}^2\right)^{\frac{1}{2}},\frac{1}{\tilde{k}}\left(1-v^2-w^2+3/\tilde{k}^2\right)^{-\frac{1}{2}} , \right.  \\
                &\left. \frac{1}{\tilde{k}}\left(1-v^2-w^2+3/\tilde{k}^2\right)^{-\frac{1}{2}}, x \right) 
	\end{aligned}
\end{equation}
where we have defined 
\begin{equation}
	\begin{aligned}
		|\mathbf{k}'-\mathbf{p}'|=u'k' \ ,  \  |\mathbf{p}'-\mathbf{q}'|=\bar{u}'p'=\bar{u}'v'k'   \ , \   q'=\bar{v}'p'=\bar{v}'v'k'~.
	\end{aligned}
\end{equation}   
In the end of this section, we calculate the third order power spectrum $\mathcal{P}^{(3)}_V$, i.e., 
\begin{equation}
	\mathcal{P}^{(3)}_V = \sum_{i,j=1}^{4}\mathcal{P}^{ij}_V(x,\tilde{k})~.\label{Z38} 
\end{equation}

\begin{figure}
    \centering
    \includegraphics[scale = 0.8]{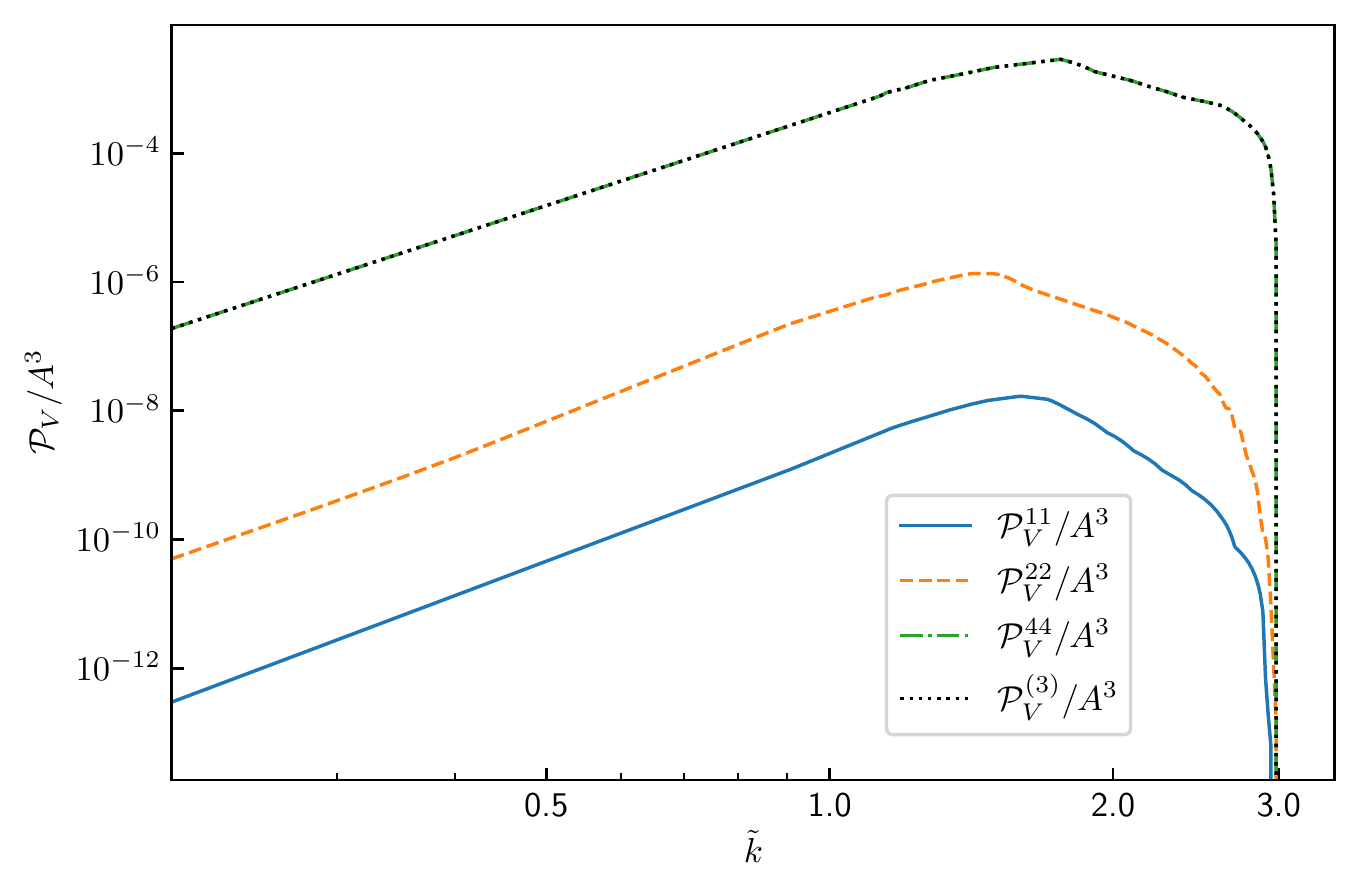}
    \caption{The power spectra of the third order vector perturbations. The black dot-dashed curve denotes the total power spectrum $\mathcal{P}_V^{(3)} \equiv \sum_{i,j=1}^4 \mathcal{P}^{ij}_V$. The diagonal terms $\mathcal{P}^{11}_V$ (blue solid curve), $\mathcal{P}^{22}_V$ (orange dashed curve) and $\mathcal{P}^{44}_V$ (green dot-dashed curve) are shown. It shows that $\mathcal{P}_V^{(3)} \simeq \mathcal{P}_V^{44}$.}\label{fig:power_spectrum}
\end{figure}

We plot the third order power spectra in Fig.~\ref{fig:power_spectrum}. It shows that the third order gravitational waves sourced by the second order scalar perturbations dominate the energy density spectrum, i.e., $\mathcal{P}_V^{(3)} \simeq \mathcal{P}_V^{44}$. This is consistent with the kernel function shown in Fig.~\ref{fig:I2_and_I3}, where $I_4^{(3)}$ have the largest amplitude for large $x$ and contributes most to the total power spectrum.

\section{Conclusions and discussions}\label{sec:4}
In this paper, we studied the second order and third order vector modes induced by first order primordial scalar perturbations. The explicit expressions of the power spectra were presented. We conclude that the second order vector modes can not be generated by a monochromatic primordial power spectrum. Moreover, we calculated the kernel functions and the power spectra of the third order vector modes for a monochromatic primordial power spectrum. It shows that the power spectra of the third order vector modes are not zero, the third order vector mode is the first non-trivial order. 

The third order vector modes have four kinds of source terms. More, precisely, the first order scalar perturbation induces the second order scalar, vector and tensor perturbations. Then, the first order scalar, the second order scalar, vector and tensor perturbations all induce the third order vector modes. The third order vector modes sourced by the second order scalar perturbations dominate the two point function $\langle V^{(3),\lambda}V^{(3),\lambda'} \rangle$ and power spectrum corresponded. And the third order vector modes sourced by the second order vector perturbations are equal to zero, this result is consistent with the conclusion of the second order vector modes.

It would be helpful to compare the third order vector modes with the third order scalar induced gravitational waves for a monochromatic primordial power spectrum \cite{Zhou:2021vcw}. First, they both have four kinds of source terms, and their total power spectra are both dominated by the source term of the second order scalar perturbations. Second, the source term of the second order vector perturbation does not affect their total power spectra. Third, the third order power spectrum of gravitational waves is composed of $\langle h^{(3),\lambda}h^{(3),\lambda'} \rangle$ and $\langle h^{(2),\lambda}h^{(4),\lambda'} \rangle$. Therefore, if we want to calculate the third order power spectrum of gravitational waves completely, we need to study the fourth order scalar induced gravitational waves $h^{(4),\lambda}$. Similarly, it is necessary to consider the correlation function $\langle V^{(2),\lambda}V^{(4),\lambda'} \rangle$ for a general primordial power spectra. 

\acknowledgments
We thank Dr. Q.H. Zhu for useful discussions. This work has been funded by the National Nature Science Foundation of China under grant No. 12075249 and 11690022, and the Key Research Program of the Chinese Academy of Sciences under Grant No. XDPB15

\appendix
\section{Decomposition Operators}\label{sec:AA}
We review the decomposition operators on FRW spacetime. The spatial tensor filed $S_{ij}$ on FRW spacetime could be decomposed as scalar, vector and tensor modes
\begin{equation}
	S_{i j}=S_{i j}^{(H)}+2 \delta_{i j} S^{(\Psi)}+2 \partial_{i} \partial_{j} S^{(E)}+\partial_{j} S_{i}^{(C)}+\partial_{i} S_{j}^{(C)} \ .
\end{equation}
We define the decomposed operators as follow
\begin{equation}
	S_{i j}^{(H)} \equiv\Lambda_{ij}^{kl}S_{k l}=\left(\mathcal{T}_{i}^{k} \mathcal{T}_{j}^{l}-\frac{1}{2} \mathcal{T}_{i j} \mathcal{T}^{k l}\right) S_{k l} \ \ , \ \ S^{(\Psi)} \equiv \frac{1}{4} \mathcal{T}^{k l} S_{k l} \ .
\end{equation}
\begin{equation}\label{De2}
	S^{(E)} \equiv \frac{1}{2} \Delta^{-1}\left(\partial^{k} \Delta^{-1} \partial^{l}-\frac{1}{2} \mathcal{T}^{k l}\right) S_{l k} \ \ , \ \ S_{i}^{(C)} \equiv \Delta^{-1} \partial^{l} \mathcal{T}_{i}^{k} S_{l k} \ .
\end{equation}
The transverse operators is defined by
\begin{equation}
	\mathcal{T}_{j}^{i} \equiv \delta_{i}^{i}-\partial^{i} \Delta^{-1} \partial_{j} \ .
\end{equation} 
These decomposed operators can only use in FRW spacetime. For arbitrary spacetime, we can't distinguish between $S_{i}^{(C)}$ and $S^{(E)}$ \cite{Nakamura:2011xy}.

\section{Higher order perturbations of Einstein equation}\label{sec:AB}
In this appendix, we use the $\texttt{xPand}$ package to study the higher order perturbations of Einstein equation on FRW spacetime \cite{Pitrou:2013hga}. The $\texttt{xPand}$ package can help us to obtain and simplify the perturbation equations. 
\subsection{Second order perturbation equation}\label{sec:AB1}
In this appendix, we study the second order perturbation equation and derive the equation of motion of second order vector mode $V^{(2)}_i$.The space-space part of the second order perturbed equation is presented as follows
\begin{equation}\label{eq:2order1}
	\begin{aligned}
		&\frac{1}{4}h^{(2)''}_{ij}+\frac{1}{2}\mathcal{H}h^{(2)'}_{ij}-\frac{1}{2}\mathcal{H}^2h^{(2)}_{ij}-\mathcal{H}'h^{(2)}_{ij}-\frac{1}{6}\kappa a^2h^{(2)}_{ij}\rho^{(0)}-\frac{4}{3}\kappa a^2 u^{(1)}_iu^{(1)}_j\rho^{(0)}-\frac{1}{4}\Delta h^{(2)}_{ij} \\
		&-\frac{1}{2}\mathcal{H}\partial_{i}V^{(2)}_j-\frac{1}{4}\partial_{i}V^{(2)'}_j-\frac{1}{2}\mathcal{H}\partial_{j}V^{(2)}_i-\frac{1}{4}\partial_{j}V^{(2)'}_i+\partial_{i}\phi^{(1)}\partial_{j}\phi^{(1)}-\partial_{i}\psi^{(1)}\partial_{j}\phi^{(1)}-\partial_{j}\psi^{(1)}\partial_{i}\phi^{(1)} \\
		&+3\partial_{i}\psi^{(1)}\partial_{j}\psi^{(1)}+2\phi^{(1)}\partial_{i}\partial_{j}\phi^{(1)}+2\psi^{(1)}\partial_{i}\partial_{j}\psi^{(1)}-\frac{1}{2}\partial_{i}\partial_{j}\phi^{(2)}+\frac{1}{2}\partial_i\partial_j\psi^{(2)}=0 \ ,
	\end{aligned}
\end{equation}
where $u^{(1)}_i$ is the first order transverse part of three dimensional velocity in energy-momentum tensor, $\rho^{(0)}$ is the zero order density in energy-momentum tensor, $\mathcal{H}'$ is the derivative of $\mathcal{H}$ with respect to $\eta$. The time-time component and space-time component of lower order perturbation equations can help us to express $u^{(1)}_i$, $\rho^{(0)}$, and $\mathcal{H}'$ in terms of $\mathcal{H}$, and the first order perturbations. Namely,
\begin{equation}\label{eq:H0}
	\begin{aligned}
		\mathcal{H}^{'}=-\mathcal{H}^2 \ \ , \ \ \rho^{(0)}=\frac{3\mathcal{H}^{2}}{\kappa a^2} \ ,
	\end{aligned}
\end{equation}
\begin{equation}\label{eq:rho1}
	\begin{aligned}
		\rho^{(1)}=\frac{-6\mathcal{H}\left(\mathcal{H}\phi^{(1)}+\psi^{(1)'}\right)+2\Delta \psi^{(1)}}{\kappa a^2} \ \ , \ \ u^{(1)}_i=-\frac{\left(\mathcal{H}\partial_i\phi^{(1)}+\partial_i\psi^{(1)'}\right)}{2\mathcal{H}^2} \ ,
	\end{aligned}
\end{equation}
Substituting Eqs.~(\ref{eq:H0})--(\ref{eq:rho1}) into Eq.~(\ref{eq:2order1}), we obtain
\begin{equation}\label{eq:2order2}
	\begin{aligned}
		&\frac{1}{4}h^{(2)''}_{ij}+\frac{1}{2}\mathcal{H}h^{(2)'}_{ij}-\frac{1}{4}\Delta h^{(2)}_{ij}-\frac{1}{2}\mathcal{H}\partial_{i}V^{(2)}_j-\frac{1}{4}\partial_{i}V^{(2)'}_j-\frac{1}{2}\mathcal{H}\partial_{j}V^{(2)}_i-\frac{1}{4}\partial_{j}V^{(2)'}_i-\frac{1}{2}\partial_{i}\partial_{j}\phi^{(2)}\\
		&+\frac{1}{2}\partial_i\partial_j\psi^{(2)}+\partial_{i}\phi^{(1)}\partial_{j}\phi^{(1)}-\partial_{i}\psi^{(1)}\partial_{j}\phi^{(1)}-\partial_{j}\psi^{(1)}\partial_{i}\phi^{(1)}-\frac{\partial_i\psi^{(1)'}\partial_{j}\phi^{(1)}}{\mathcal{H}}-\frac{\partial_i\phi^{(1)}\partial_j\psi^{(1)'}}{\mathcal{H}} \\
		&-\frac{\partial_i\psi^{(1)'}\partial_j\psi^{(1)'}}{\mathcal{H}^2}+3\partial_{i}\psi^{(1)}\partial_{j}\psi^{(1)}+2\phi^{(1)}\partial_{i}\partial_{j}\phi^{(1)}+2\psi^{(1)}\partial_{i}\partial_{j}\psi^{(1)}=0 \ .
	\end{aligned}
\end{equation}
Using the decomposition operator $\Delta^{-1} \partial^{l} \mathcal{T}_{i}^{k}$ in Eq.~(\ref{De2}) to extract the equation of motion of second order vector perturbation, we obtain
\begin{equation}\label{eq:eV}
	\begin{aligned}
		V_{l}^{(2)'}(\eta,\mathbf{x})+2 \mathcal{H} V_{l}^{(2)}(\eta,\mathbf{x}) =-4 \Delta^{-1} \mathcal{T}_{l}^{r} \partial^{s} \mathcal{S}^{(2)}_{rs}(\eta,\mathbf{x}) \ ,
	\end{aligned}
\end{equation}
where the source term is given by
\begin{equation}
	\begin{aligned}
		\Delta^{-1} \mathcal{T}_{l}^{r} \partial^{s} \mathcal{S}^{(2)}_{rs}(\eta,\mathbf{x})=& \Delta^{-1} \mathcal{T}_{l}^{r} \partial^{s}\Bigg(\partial_{r} \phi^{(1)} \partial_{s} \phi^{(1)}-\frac{1}{ \mathcal{H}}\left(\partial_{r} \phi^{(1)'} \partial_{s} \phi^{(1)}+\partial_{r} \phi^{(1)}  \partial_{s} \phi^{(1)'}\right)\\
		&+4 \phi^{(1)} \partial_{r} \partial_{s} \phi^{(1)}-\frac{1}{ \mathcal{H}^{2}} \partial_{r} \phi^{(1)'} \partial_{s}  \phi^{(1)'}\Bigg) \ .
	\end{aligned}
\end{equation}
Here, we have set $\psi^{(1)}=\phi^{(1)}$ from the equation of motion of first order scalar modes.
\subsection{Third order perturbation equation}\label{sec:AB2}
In this appendix, we investigate the third order perturbation equation and obtain the equation of motion of third order vector mode $V^{(3)}_i$. The space-space part of the third order perturbed equation is presented as follows,
\begin{equation}\label{eq:eq3}
	\begin{aligned}
		&-\frac{1}{6}\mathcal{H}\partial_{i}V^{(3)}_j-\frac{1}{6}\mathcal{H}\partial_{j}V^{(3)}_i-\frac{1}{12}\partial_{i}V^{(3)'}_j-\frac{1}{12}\partial_{j}V^{(3)'}_i+\frac{1}{6}\partial_{i}\partial_{j}\psi^{(3)}-\frac{1}{6}\partial_{i}\partial_{j}\phi^{(3)}\\
		&+\frac{1}{12}h^{(3)''}_{ij}+\frac{1}{6}h^{(3)'}_{ij}\mathcal{H}-\frac{1}{6}h^{(3)}_{ij}\mathcal{H}^2-\frac{1}{3}h^{(3)}_{ij}\mathcal{H}^{'}-\frac{1}{18}\kappa a^2h^{(3)}_{ij}\rho^{(0)}-\frac{1}{12}\Delta h^{(3)}_{ij}-\frac{2}{3}\kappa a^2 V_j^{(2)}u^{(1)}_i\rho^{(0)}\\
		&-\frac{2}{3}\kappa a^2 V_i^{(2)}u^{(1)}_j\rho^{(0)}-\frac{2}{3}\kappa a^2u^{(1)}_iu^{(2)}_j\rho^{(0)}-\frac{2}{3}\kappa a^2u^{(1)}_ju^{(2)}_i\rho^{(0)}-\frac{1}{6}\kappa a^2h^{(2)}_{ij}\rho^{(1)}-\frac{4}{3}\kappa a^2u^{(1)}_iu^{(1)}_j\rho^{(1)}\\
		&-h^{(2)'}_{ij}\mathcal{H}\phi^{(1)}+h^{(2)}_{ij}\mathcal{H}^2\phi^{(1)}+2h^{(2)}_{ij}\mathcal{H}^{'}\phi^{(1)}+\frac{16}{3}\kappa a^2u^{(1)}_iu^{(1)}_j\rho^{(0)}\phi^{(1)}+4h^{(2)}_{ij}\mathcal{H}\phi^{(1)'}+\frac{3}{2}h^{(2)}_{ij}\phi^{(1)''}\\
		&-\frac{1}{2}h^{(2)}_{ij}\Delta \phi^{(1)}+\frac{1}{2}h^{b(2)}_{i}\partial_b\partial_j\phi^{(1)}+\frac{1}{2}h^{b(2)}_{j}\partial_b\partial_i\phi^{(1)}-\partial_bh^{(2)}_{ij}\partial^b\phi^{(1)}+\mathcal{H}\phi^{(1)}\partial_iV^{(2)}_j+\frac{1}{2}\phi^{(1)'}\partial_iV^{(2)}_j\\
		&+\frac{1}{2}\partial^b\phi^{(1)}\partial_ih^{(2)}_{jb}-\frac{1}{2}V^{(2)'}_j\partial_i\phi^{(1)}-V^{(2)}_j\mathcal{H}\partial_i\phi^{(1)}-\frac{1}{2}V^{(2)}_j\partial_i\phi^{(1)'}+\mathcal{H}\phi^{(1)}\partial_jV^{(2)}_i+\frac{1}{2}\phi^{(1)'}\partial_jV^{(2)}_i\\
		&+\frac{1}{2}\phi^{(1)}\partial_jV^{(2)}_i+\frac{1}{2}\partial^b\phi^{(1)}\partial_jh^{(2)}_{ib}-\frac{1}{2}V^{(2)'}_i\partial_j\phi^{(1)}-V^{(2)}_i\mathcal{H}\partial_j\phi^{(1)}-\frac{1}{2}V^{(2)}_i\partial_j\phi^{(1)'}+8\phi^{(1)}\partial_i\phi^{(1)}\partial_j\phi^{(1)}\\
		&+\partial_i\psi^{(2)}\partial_j\phi^{(1)}+\partial_i\phi^{(1)}\partial_j\psi^{(2)}+\phi^{(2)}\partial_i\partial_j\phi^{(1)}+\phi^{(1)}\partial_i\partial_j\phi^{(2)}+\psi^{(2)}\partial_i\partial_j\phi^{(1)}+\phi^{(1)}\partial_i\partial_j\psi^{(2)} \\
		&-\frac{1}{2}\phi^{(1)}\Delta h^{(2)}_{ij}-\frac{1}{2}h^{(2)''}_{ij}\phi^{(1)}+\frac{1}{2}\phi^{(1)}\partial_iV^{(2)'}_j=0
		\ ,
	\end{aligned}
\end{equation}
where $u^{(1)}_i$ and $u^{(2)}_i$ are first and second order transverse part of three dimensional velocity in  energy-momentum tensor. We have set $\psi^{(1)}=\phi^{(1)}$ from the equation of motion of first order scalar perturbations. We express $\rho^{(0)}$, $\rho^{(1)}$, $u^{(1)}$, $u^{(2)}$, and $\mathcal{H}^{'}$ in Eq.~(\ref{eq:eq3}) in terms of $\mathcal{H}$, the first order perturbations, and the second order perturbations,
\begin{equation}\label{eq:e1}
	\begin{aligned}
		\mathcal{H}^{'}=-\mathcal{H}^2 \ \ , \ \ \rho^{(0)}=\frac{3\mathcal{H}^{2}}{\kappa a^2} \ ,
	\end{aligned}
\end{equation}
\begin{equation}\label{eq:e2}
	\begin{aligned}
		\rho^{(1)}=\frac{-6\mathcal{H}\left(\mathcal{H}\phi^{(1)}+\psi^{(1)'}\right)+2\Delta \psi^{(1)}}{\kappa a^2} \ \ , \ \ u^{(1)}_i=-\frac{\left(\mathcal{H}\partial_i\phi^{(1)}+\partial_i\psi^{(1)'}\right)}{2\mathcal{H}^2} \ ,
	\end{aligned}
\end{equation}
\begin{equation}\label{eq:e3}
	\begin{aligned}
		\rho^{(2)}&=\frac{1}{4\kappa a^2\mathcal{H}}\Bigg[ 2\mathcal{H}\Bigg( 12\mathcal{H}^3\left( 4(\phi^{(1)})^2-\phi^{(2)}\right)-12\mathcal{H}^2\psi^{(2)'}-8\partial_b\phi^{(1)'} \partial^b\phi^{(1)}\\
		&+\mathcal{H}\left(12(\phi^{(1)'})^2+32\phi^{(1)}\Delta\phi^{(1)}+4\Delta\psi^{(2)}+8\partial_b\phi^{(1)}\partial^b\phi^{(1)}\right)   \Bigg)-8\partial_b\phi^{(1)'}\partial^b\phi^{(1)'}\Bigg] \ ,
	\end{aligned}
\end{equation}
\begin{equation}\label{eq:e4}
	\begin{aligned}
		u^{(2)}_i&=-V^{(2)}_i+\frac{1}{32\mathcal{H}^4}\Bigg[\frac{64}{3}\Delta\phi^{(1)}\partial_i\phi^{(1)'}-\frac{64}{3}\mathcal{H}\left(-\Delta \phi^{(1)}\partial_i\phi^{(1)}+3\phi^{(1)'}\partial_i\phi^{(1)'} \right)\\ 
		&+3\mathcal{H}^2\left(\frac{4}{3}\Delta V_i^{(2)}-32\phi^{(1)'}\partial_i\phi^{(1)}-\frac{160}{3} \phi^{(1)}\partial_i\phi^{(1)'}-\frac{16}{3}\partial_i\psi^{(2)'}\right)\\
		&+12\mathcal{H}^3\left(-\frac{8}{3}\phi^{(1)}\partial_i\phi^{(1)}-\frac{4}{3}\partial_i\phi^{(2)}\right)\Bigg] \ .
	\end{aligned}
\end{equation}
Substituting Eqs.~(\ref{eq:e1})--(\ref{eq:e4}) into Eq.~(\ref{eq:eq3}), we obtain
\begin{equation}\label{eq:eq31}
	\begin{aligned}
		&+\frac{2}{3\mathcal{H}^2}\Delta\phi^{(1)}\partial_l\phi^{(1)}\partial_m\phi^{(1)}+\frac{2}{3\mathcal{H}^4}\Delta\phi^{(1)}\partial_l\phi^{(1)'}\partial_m\phi^{(1)'}-\frac{3}{\mathcal{H}^2}\phi^{(1)'}\partial_l\phi^{(1)'}\partial_m\phi^{(1)}-\frac{3}{\mathcal{H}^2}\phi^{(1)'}\partial_m\phi^{(1)'}\partial_l\phi^{(1)}\\
		&+\frac{2}{3\mathcal{H}^3}\Delta\phi^{(1)}\partial_l\phi^{(1)'}\partial_m\phi^{(1)}+\frac{2}{3\mathcal{H}^3}\Delta\phi^{(1)}\partial_m\phi^{(1)'}\partial_l\phi^{(1)}-\frac{2}{\mathcal{H}^3}\phi^{(1)'}\partial_l\phi^{(1)'}\partial_m\phi^{(1)'}-\frac{4}{\mathcal{H}^2}\phi^{(1)}\partial_l\phi^{(1)'}\partial_m\phi^{(1)'} \\
		&-\frac{1}{6}\mathcal{H}\partial_{l}V^{(3)}_m-\frac{1}{6}\mathcal{H}\partial_{m}V^{(3)}_l-\frac{1}{12}\partial_{l}V^{(3)'}_m-\frac{1}{12}\partial_{m}V^{(3)'}_l+\frac{1}{6}\partial_{l}\partial_{m}\psi^{(3)}-\frac{1}{6}\partial_{l}\partial_{m}\phi^{(3)}\\
		&+h_{lm}^{(3)''}(\eta,\mathbf{x})+2 \mathcal{H}  h_{i lm}^{(3)'}(\eta,\mathbf{x})-\Delta h_{lm}^{(3)}(\eta,\mathbf{x})+12\phi^{(1)}\partial_l\phi^{(1)}\partial_m\phi^{(1)}-\frac{4}{\mathcal{H}}\phi^{(1)'}\partial_l\phi^{(1)}\partial_m\phi^{(1)}\\
		&-\frac{1}{2}\phi^{(1)}\left( h_{lm}^{(2)''}+2 \mathcal{H}  h_{lm}^{(2)'}-\Delta h_{lm}^{(2)}\right)-\phi^{(1)}\Delta h_{lm}^{(2)}-\phi^{(1)'}\mathcal{H}h_{lm}^{(2)}-\frac{1}{3}\Delta \phi^{(1)}h_{lm}^{(2)}-\partial^b \phi^{(1)}\partial_b h_{lm}^{(2)} \\
		&+\phi^{(1)}\partial_l\left(V_m^{(2)'}+2 \mathcal{H}V_m^{(2)} \right)+\phi^{(1)}\partial_m\left(V_l^{(2)'}+2 \mathcal{H}V_l^{(2)} \right)+\phi^{(1)'}\left(\partial_lV_m^{(2)}+\partial_mV_l^{(2)}\right)\\
		&-\frac{\phi^{(1)}}{8\mathcal{H}}\left(\partial_m\Delta V_l^{(2)}+\partial_l\Delta V_m^{(2)}\right)-\frac{\phi^{(1)'}}{8\mathcal{H}^2}\left(\partial_m\Delta V_l^{(2)}+\partial_l\Delta V_m^{(2)}\right)+\frac{1}{\mathcal{H}}\left(\phi^{(1)}\partial_l\partial_m\psi^{(2)'}\right) \\
		&+\frac{1}{\mathcal{H}}\left(\phi^{(1)'}\partial_l\partial_m\phi^{(2)}\right)+\frac{1}{\mathcal{H}^2}\left(\phi^{(1)'}\partial_l\partial_m\psi^{(2)'}\right)+3\left(\phi^{(1)}\partial_l\partial_m\phi^{(2)}\right)=0 \ .
	\end{aligned}
\end{equation}
Using the decomposition operator $\Delta^{-1} \partial^{l} \mathcal{T}_{i}^{k}$ in Eq.~(\ref{De2}), we can extract the equation of motion of third vector perturbation.

\section{The correlation function}
In the appendix, we simplify the four-point correlation function and the six-point correlation function in the power spectra of vector mode.

\subsection{Four-point correlation function}\label{sec:C4}
The four-point function can be simplified in terms of Wick's theorem.
\begin{equation}\label{eq:Wick2}
\begin{aligned}
	\left\langle\Phi_{\mathbf{k}-\mathbf{p}} \Phi_{\mathbf{p}} \Phi_{\mathbf{k}^{\prime}-\mathbf{p}^{\prime}} \Phi_{\mathbf{p}^{\prime}}\right\rangle &=\left\langle\Phi_{\mathbf{k}-\mathbf{p}} \Phi_{\mathbf{p}}\right\rangle\left\langle\Phi_{\mathbf{k}^{\prime}-\mathbf{p}^{\prime}} \Phi_{\mathbf{p}^{\prime}}\right\rangle+\left\langle\Phi_{\mathbf{k}-\mathbf{p}} \Phi_{\mathbf{k}^{\prime}-\mathbf{p}^{\prime}}\right\rangle\left\langle\Phi_{\mathbf{p}} \Phi_{\mathbf{p}^{\prime}}\right\rangle+\left\langle\Phi_{\mathbf{k}-\mathbf{p}} \Phi_{\mathbf{p}^{\prime}}\right\rangle\left\langle\Phi_{\mathbf{p}} \Phi_{\mathbf{k}^{\prime}-\mathbf{p}^{\prime}}\right\rangle \\
	&=\frac{\left(2 \pi^{2}\right)^{2}}{|k-p|^{3}\left|k^{\prime}-q\right|^{3}} \delta(\mathbf{k}) \delta\left(\mathbf{k}^{\prime}\right) \mathcal{P}_{\Phi}(k-p) \mathcal{P}_{\Phi}\left(k^{\prime}-p^{\prime}\right) \\
	&+\frac{\left(2 \pi^{2}\right)^{2}}{|k-p|^{3} p^{3}} \delta\left(\mathbf{k}-\mathbf{p}+\mathbf{k}^{\prime}-\mathbf{p}^{\prime}\right) \delta\left(\mathbf{p}+\mathbf{p}^{\prime}\right) \mathcal{P}_{\Phi}(k-p) \mathcal{P}_{\Phi}(p) \\
	&+\frac{\left(2 \pi^{2}\right)^{2}}{|k-p|^{3} p^{3}} \delta\left(\mathbf{k}-\mathbf{p}+\mathbf{p}^{\prime}\right) \delta\left(\mathbf{p}+\mathbf{k}^{\prime}-\mathbf{p}^{\prime}\right) \mathcal{P}_{\Phi}(k-p) \mathcal{P}_{\Phi}(p) \\
	&=\frac{\left(2 \pi^{2}\right)^{2}}{|k-p|^{3} p^{3}} \delta\left(\mathbf{k}+\mathbf{k}^{\prime}\right)\left(\delta\left(\mathbf{p}+\mathbf{p}^{\prime}\right)+\delta\left(\mathbf{k}-\mathbf{p}+\mathbf{p}^{\prime}\right)\right) \mathcal{P}_{\Phi}(k-p) \mathcal{P}_{\Phi}(p) \ ,
\end{aligned}
\end{equation}
where in the last line we have neglected the first term which corresponds to the bubble diagram $\int d^3p  \mathcal{P}_{\Phi}(p) $.

\subsection{Six-point correlation function}\label{sec:C6}
In this appendix, we use Wick's theorem to simplify the six-point correlation function in Sec.~\ref{sec:3.2}. We define $\mathcal{C}(\mathbf{k},\mathbf{k}',\mathbf{p},\mathbf{p}',\mathbf{q},\mathbf{q}')$ as
\begin{equation}\label{eq:C}
	\begin{aligned}
		&\mathcal{C}(\mathbf{k},\mathbf{k}',\mathbf{p},\mathbf{p}',\mathbf{q},\mathbf{q}')=\frac{1}{(2\pi^2)^3}\int{\rm d}^3\mathbf{k}' \langle \Phi_{\mathbf{k}-\mathbf{p}} \Phi_{\mathbf{p}-\mathbf{q}} \Phi_{\mathbf{q}}\Phi_{\mathbf{k}'-\mathbf{p}'} \Phi_{\mathbf{p}'-\mathbf{q}'} \Phi_{\mathbf{q}'} \rangle \\
		&=\Bigg[\left( \frac{1}{(k-p)^3(p-q)^3q^3}\delta(\mathbf{p}+\mathbf{p}')\delta(\mathbf{q}+\mathbf{q}')P_{\Phi}(k-p)P_{\Phi}(p-q)P_{\Phi}(q)\right) \\
		&+\left( \frac{1}{(k-p)^3(p-q)^3q^3}\delta(\mathbf{p}+\mathbf{p}')\delta(\mathbf{q}+\mathbf{p}'-\mathbf{q}')P_{\Phi}(k-p)P_{\Phi}(p-q)P_{\Phi}(q)\right) \\
		&+\left( \frac{1}{(k-p)^3(p-q)^3q^3}\delta(\mathbf{p}-\mathbf{q}-\mathbf{k}-\mathbf{p}')\delta(\mathbf{q}+\mathbf{q}')P_{\Phi}(k-p)P_{\Phi}(p-q)P_{\Phi}(q)\right) \\
		&+\left( \frac{1}{(k-p)^3(p-q)^3q^3}\delta(\mathbf{p}-\mathbf{q}+\mathbf{q}')\delta(\mathbf{q}-\mathbf{k}-\mathbf{p}')P_{\Phi}(k-p)P_{\Phi}(p-q)P_{\Phi}(q)\right) \\
		&+\left( \frac{1}{(k-p)^3(p-q)^3q^3}\delta(\mathbf{p}-\mathbf{k}-\mathbf{q}')\delta(\mathbf{q}+\mathbf{p}'-\mathbf{q}')P_{\Phi}(k-p)P_{\Phi}(p-q)P_{\Phi}(q)\right) \\
		&+\left( \frac{1}{(k-p)^3(p-q)^3q^3}\delta(-\mathbf{k}+\mathbf{p}-\mathbf{q}')\delta(\mathbf{q}-\mathbf{k}-\mathbf{p}')P_{\Phi}(k-p)P_{\Phi}(p-q)P_{\Phi}(q)\right)\Bigg] \ .
	\end{aligned}
\end{equation}
Here, we have neglected the bubble diagram in the Wick's expansion.

\section{The power spectra of vector modes}\label{sec:D}
In this appendix, we calculate the power spectra of second order and third order vector modes. The explicit expressions of the power spectra will be given in this appendix.

\subsection{The power spectrum of second order vector mode}\label{sec:D1}
Substituting Eq.~(\ref{eq:Wick2}) into Eq.~(\ref{eq:Power2}) and integrating over $\mathbf{k}'$ and $\mathbf{p}'$, we obtain
\begin{equation}
	\begin{aligned}
		P_V^{(2)}=\sum_{a=1}^{2} \mathcal{P}_a^{(2)}= \mathcal{P}_1^{(2)}+ \mathcal{P}_2^{(2)}\ ,
	\end{aligned}
\end{equation}
where the summation of $a$ comes from the Wick's expression in the last line of Eq.~(\ref{eq:Wick2}). More precisely, substituting the first term in the last line of Eq.~(\ref{eq:Wick2}) into Eq.~(\ref{eq:Power2}) and integrating over $\mathbf{k}'$ and $\mathbf{p}'$, we obtain
\begin{equation}
	\begin{aligned}
		 \mathcal{P}_1^{(2)}&=k^3\pi^2\int\frac{d^3p}{(2\pi)^3}\frac{1}{|k-p|^3p^3}\delta_{\lambda \lambda'}\frac{k^se^{\lambda,r}(\mathbf{k})}{k^2}\frac{k^le^{\lambda',m}(\mathbf{k})}{k^2}p_rp_sp_lp_m \left(I^{(2)}_V\left(|k-p|,p,\eta  \right)\right)^2\mathcal{P}_{\Phi}(k-p) \mathcal{P}_{\Phi}(p) \\
		&=k^3\pi^2\int\frac{d^3p}{(2\pi)^3}\frac{1}{|k-p|^3p^3}\frac{k^sk^l\left(\delta^{mr}-\frac{k^mk^r}{k^2}\right)}{k^4}p_rp_sp_lp_m \left(I^{(2)}_V\left(|k-p|,p,\eta  \right)\right)^2\mathcal{P}_{\Phi}(k-p) \mathcal{P}_{\Phi}(p)\\
		&=\pi^{2} k^{3} \int \frac{p^{2} \mathrm{~d} p \sin \theta \mathrm{d} \theta \mathrm{d} \phi}{(2 \pi)^{3}}\left\{\frac{1}{|k-p|^{3} p^{3}} \left(\frac{(k\cdot p)^2(p^2-\frac{(k\cdot p)^2}{k^2})}{k^4}\right) \left(I^{(2)}_V\left(|k-p|,p,\eta  \right)\right)^2 \mathcal{P}_{\Phi}(k-p) \mathcal{P}_{\Phi}(p)\right\} \\
	    &=-\frac{1}{4} \int \mathrm{d} p \mathrm{~d} \cos \theta\left\{\frac{k^{3}}{|k-p|^{3} p}\left(\frac{(k\cdot p)^2(p^2-\frac{(k\cdot p)^2}{k^2})}{k^4}\right) \left(I^{(2)}_V\left(|k-p|,p,\eta  \right)\right)^2 \mathcal{P}_{\Phi}(k-p) \mathcal{P}_{\Phi}(p)\right\} \\
	    &=\frac{1}{4} \int_{0}^{\infty} \mathrm{d} v \int_{|1-v|}^{1+v} \mathrm{~d} u\left(\frac{\left( 1+v^2-u^2 \right)\left(4 v^{2}-\left(1+v^{2}-u^{2}\right)^{2}\right)}{16u^2v^2}\right)\left(kI^{(2)}_V\left(|k-p|,p,\eta  \right)\right)^2  \mathcal{P}_{\Phi}(k u) \mathcal{P}_{\Phi}(k v) \ ,
	\end{aligned}
\end{equation}
where in the second line we have used the property of the polarization vectors  $\sum_{\lambda}	e^{\lambda,m}(\mathbf{k})e^{\lambda,r}(\mathbf{k})+\frac{k^mk^r}{k^2}=\delta^{mr}$. In the last line, we have used the relations  $|\mathbf{k}-\mathbf{p}|=uk$, $|\mathbf{p}|=vk$, and $\cos\theta=\frac{1+v^2-u^2}{2v}$.
Substituting the second term in the last line of Eq.~(\ref{eq:Wick2}) into Eq.~(\ref{eq:Power2}) and integrating over $\mathbf{k}'$ and $\mathbf{p}'$, we obtain
\begin{equation}
	\begin{aligned}
		\mathcal{P}_2^{(2)}&=k^3\pi^2\int\frac{d^3p}{(2\pi)^3}\frac{1}{|k-p|^3p^3}\delta_{\lambda \lambda'}\frac{k^se^{\lambda,r}(\mathbf{k})}{k^2}\frac{k^le^{\lambda',m}(\mathbf{k})}{k^2}p_rp_s(p_l-k_l)(p_m-k_m)  \\
		&\times I^{(2)}_V\left(|k-p|,p,\eta  \right)I^{(2)}_V\left(p,|k-p|,\eta  \right)  \mathcal{P}_{\Phi}(k-p) \mathcal{P}_{\Phi}(p) \\
		&=k^3\pi^2\int\frac{d^3p}{(2\pi)^3}\frac{1}{|k-p|^3p^3}\frac{\left( k\cdot p \right)\left( k\cdot p-k^2 \right)(p^2-\frac{(k\cdot p)^2}{k^2})}{k^4}  \\
		&\times I^{(2)}_V\left(|k-p|,p,\eta  \right)I^{(2)}_V\left(p,|k-p|,\eta  \right)  \mathcal{P}_{\Phi}(k-p) \mathcal{P}_{\Phi}(p) \\
		&=\frac{1}{4} \int_{0}^{\infty} \mathrm{d} v \int_{|1-v|}^{1+v} \mathrm{~d} u\left(\frac{\left(\left( 1+v^2-u^2 \right)^2-2\left(1+v^2-u^2\right)\right)\left(4 v^{2}-\left(1+v^{2}-u^{2}\right)^{2}\right)}{16v^2u^2}\right) \\
		&\times k^2 I^{(2)}_V\left(|k-p|,p,\eta  \right)I^{(2)}_V\left(p,|k-p|,\eta  \right)  \mathcal{P}_{\Phi}(k u) \mathcal{P}_{\Phi}(k v) \ .
	\end{aligned}
\end{equation}

\subsection{The power spectrum of third order vector mode}\label{sec:D2}
The calculations of the power spectrum of third order vector mode are similar to the second order power spectrum. The integral and measure in Eq.~(\ref{eq:Ppij}) are much more difficult to calculate, compared with the second order power spectrum. As we mentioned in Sec.~\ref{sec:3.2}, the variables $\mathbf{p}'$ and $\mathbf{q}'$ can be integrated in terms of the three dimensional delta functions in Eq.~(\ref{eq:C}). Then, the integral and measure in Eq.~(\ref{eq:Ppij}) in spherical coordinate system can be written as
\begin{equation}\label{eq:i1}
	\begin{aligned}
		\int \mathrm{d}^3p \int \mathrm{d}^3q=
		\int p^2 \mathrm{d}p \mathrm{d}\left(\cos \theta_{p} \right)\mathrm{d}\phi_{p}	\int q^2 \mathrm{d}q \mathrm{d}\left(\cos \theta_{q} \right)\mathrm{d}\phi_{q} \ .
	\end{aligned}
\end{equation}
We set $\mathbf{k}/|\mathbf{k}|$=$\left( 0,0,1 \right)$, then $\cos \theta_{p}$, $\cos \theta_{q}$, $\cos\phi_{p}$, and $\phi_{q}$ in Eq.~(\ref{eq:i1}) take the form of 
\begin{equation}\label{eq:i2}
	\begin{aligned}
		\cos \phi_{p}&=X=\frac{\left(\mathbf{p}-\frac{1}{k^{2}} \mathbf{k}(\mathbf{k} \cdot \mathbf{p})\right) \cdot\left(\mathbf{q}-\frac{1}{k^{2}} \mathbf{k}(\mathbf{k} \cdot \mathbf{q})\right)}{\left|\mathbf{p}-\frac{1}{k^{2}} \mathbf{k}(\mathbf{k} \cdot \mathbf{p})\right|\left|\mathbf{q}-\frac{1}{k^{2}} \mathbf{k}(\mathbf{k} \cdot \mathbf{q})\right|} \\
		&=\frac{\mathbf{p} \cdot \mathbf{q}-\frac{1}{k^{2}}(\mathbf{k} \cdot \mathbf{p})(\mathbf{k} \cdot \mathbf{q})}{\sqrt{\left(p^{2}-\frac{1}{k^{2}}(\mathbf{k} \cdot \mathbf{p})^{2}\right)\left(q^{2}-\frac{1}{k^{2}}(\mathbf{k} \cdot \mathbf{q})^{2}\right)}} \\
		&=\frac{\frac{k^{2} v^{2}\left(1+\bar{v}^{2}-\bar{u}^{2}\right)}{2}-\frac{1}{k^{2}}\left(\frac{k^{2}\left(1+v^{2}-u^{2}\right)}{2}\right)\left(\frac{k^{2}\left(1+\bar{v}^{2} v^{2}-w^{2}\right)}{2}\right)}{\sqrt{\left(v^{2} k^{2}-\frac{1}{k^{2}}\left(\frac{k^{2}\left(1+v^{2}-u^{2}\right)}{2}\right)^{2}\right)\left((\bar{v} v k)^{2}-\frac{1}{k^{2}}\left(\frac{k^{2}\left(1+\bar{v}^{2} v^{2}-w^{2}\right)}{2}\right)^{2}\right)}} \\
		&=\frac{-1+u^{2}+v^{2}-2 \bar{u}^{2} v^{2}+v^{2} \bar{v}^{2}+u^{2} v^{2} \bar{v}^{2}-v^{4} \bar{v}^{2}+w^{2}-u^{2} w^{2}+v^{2} w^{2}}{\sqrt{\left(1-2 u^{2}+u^{4}-2 v^{2}-2 u^{2} v^{2}+v^{4}\right)\left(1-2 v^{2} \bar{v}^{2}+v^{4} \bar{v}^{4}-2 w^{2}-2 v^{2} \bar{v}^{2} w^{2}+w^{4}\right)}}  \ ,
	\end{aligned}
\end{equation}

\begin{equation}\label{eq:i3}
	\begin{aligned}
		\cos\theta_p=\frac{|\mathbf{k}-\mathbf{p}|^2-k^2-p^2}{-2pk}=\frac{1+v^2-u^2}{2v}  \ ,
	\end{aligned}
\end{equation}

\begin{equation}\label{eq:i4}
	\begin{aligned}
		\cos\theta_q=\frac{|\mathbf{k}-\mathbf{q}|^2-k^2-q^2}{-2qk}=\frac{1+(\bar{v}v)^2-w^2}{2\bar{v}v}  \ ,
	\end{aligned}
\end{equation}

\begin{equation}\label{eq:i5}
	\begin{aligned}
		\int \mathrm{d}\phi_q=2\pi \ ,
	\end{aligned}
\end{equation}
where the variables $u$, $v$ , $\bar{u}$, $\bar{v}$, and $w$ are defined in Sec.~\ref{sec:3.1}. Substituting Eq.~(\ref{eq:i2})--Eq.~(\ref{eq:i5}) and Eq.~(\ref{eq:C}) into Eq.~(\ref{eq:Ppij}), we can obtain Eq.~(\ref{eq:Pex}).

\section{The expressions of $\mathbb{P}^{ij}(\eta, \mathbf{k})$}\label{sec:E}
Explicit expressions of $\mathbb{P}^{ij}_a$ in Eq.~(\ref{Z3}) are shown as follows, 

\begin{equation}
	\begin{aligned}
		&\mathbb{P}^{11}_1=\mathbb{P}^{11}	\left( \textbf{k}, \textbf{p}, - \textbf{p}, \textbf{q}, - \textbf{q}	\right) +\mathbb{P}^{11} \left( \textbf{k}, \textbf{p}, -	\textbf{p}, \textbf{q}, \textbf{q} - \textbf{p} \right) \\
		&=-\frac{k^2}{8\tilde{k}^8} \left(\text{ks}^2 \left(v^2+w^2\right)-2\right) \left(\text{ks}^6 w^2 \left(w^2-1\right) \left(v^2+w^2-1\right)-\text{ks}^4 \left(v^2 \left(2 w^2+1\right)+4 w^4-2 w^2\right) \right. \\
		&\left.+\text{ks}^2 \left(v^2+5 w^2+2\right)-2\right) \ ,
	\end{aligned}
\end{equation}

\begin{equation}
	\begin{aligned}
		&\mathbb{P}^{11}_2=\mathbb{P}^{11} \left( \textbf{k}, \textbf{p}, \textbf{q} -	\textbf{k}, \textbf{q}, \textbf{q} - \textbf{p} \right) +\mathbb{P}^{11} \left( \textbf{k}, \textbf{p}, \textbf{q} - \textbf{k},	\textbf{q}, - \textbf{k} + \textbf{p} \right)\\
		&=-\frac{k^2 \left(\text{ks}^2 \left(w^2+1\right)-1\right) \left(\text{ks}^2 \left(v^2+w^2\right)-2\right) \left(\text{ks}^4 v^2 \left(v^2+w^2-1\right)-\text{ks}^2 \left(3 v^2+w^2-2\right)+2\right)}{8 \text{ks}^8} \ ,
	\end{aligned}
\end{equation}

\begin{equation}
	\begin{aligned}
		&\mathbb{P}^{11}_3=\mathbb{P}^{11} \left( \textbf{k}, \textbf{p}, \textbf{p}-\textbf{q} -	\textbf{k}, \textbf{q}, -\textbf{q} \right) +\mathbb{P}^{11} \left( \textbf{k}, \textbf{p}, \textbf{p}-\textbf{q} - \textbf{k},	\textbf{q}, \textbf{p}- \textbf{k}   \right)\\
		&=\frac{k^2 \left(\text{ks}^2 \left(v^2+w^2\right)-2\right) \left(\text{ks}^6 v^2 w^2 \left(w^2-1\right)-\text{ks}^4 \left(v^2 \left(2 w^2+1\right)+w^4-2 w^2-1\right)+\text{ks}^2 \left(v^2+2 w^2\right)-1\right)}{8 \text{ks}^8} \ ,
	\end{aligned}
\end{equation}

\begin{equation}
	\begin{aligned}
		&\mathbb{P}^{22}_1=\mathbb{P}^{22}	\left( \textbf{k}, \textbf{p}, - \textbf{p}, \textbf{q}, - \textbf{q}	\right) +\mathbb{P}^{22} \left( \textbf{k}, \textbf{p}, -	\textbf{p}, \textbf{q}, \textbf{q} - \textbf{p} \right) \\
		&=\frac{k^2}{512 \text{ks}^{12} v^4} \left(-4 \text{ks}^4 v^2 \left(\text{ks}^2 v^2-4\right) \left(\text{ks}^6 v^2 \left(3 v^4+2 v^2 \left(8 w^2-5\right)+16 w^4-16 w^2+3\right) \right.\right.  \\
		&\left.\left.+\text{ks}^4 \left(-18 v^4+v^2 \left(34-48 w^2\right)+4\right)+\text{ks}^2 \left(27 v^2-8\right)+4\right)-\left(\text{ks}^6 v^2 \left(v^4+v^2 \left(8 w^2-6\right)+8 w^4-8 w^2+1\right) \right.\right. \\
		&\left.\left.+\text{ks}^4 \left(-6 v^4+v^2 \left(22-24 w^2\right)+4\right)+\text{ks}^2 \left(9 v^2-8\right)+4\right)^2\right)\ ,
	\end{aligned}
\end{equation}

\begin{equation}
	\begin{aligned}
		&\mathbb{P}^{22}_2=\mathbb{P}^{22} \left( \textbf{k}, \textbf{p}, \textbf{q} -	\textbf{k}, \textbf{q}, \textbf{q} - \textbf{p} \right) +\mathbb{P}^{22} \left( \textbf{k}, \textbf{p}, \textbf{q} - \textbf{k},	\textbf{q}, - \textbf{k} + \textbf{p} \right)\\
		&=-\frac{k^2}{512 \text{ks}^{12} v^2 w^2} \left(\text{ks}^{12} v^2 w^2 \left(8 v^8+8 v^6 \left(9 w^2-1\right)+v^4 \left(129 w^4-62 w^2-7\right)+2 v^2 \left(36 w^6-31 w^4-7 w^2+4\right)\right. \right. \\
		&\left.\left.+8 w^8-8 w^6-7 w^4+8 w^2-1\right)+\text{ks}^{10} \left(-72 v^8 w^2+v^6 \left(-438 w^4+38 w^2+4\right)+v^4 \left(-438 w^6+170 w^4 \right.\right.\right. \\
		&\left.\left.\left.+74 w^2\right)+v^2 \left(-72 w^8+38 w^6+74 w^4+2 w^2-4\right)+4 w^2 \left(w^4-1\right)\right)+\text{ks}^8 \left(v^6 \left(225 w^2-8\right)\right.\right. \\
		&\left.\left.+12 v^4 \left(63 w^4-11 w^2-2\right)+v^2 \left(225 w^6-132 w^4-144 w^2+8\right)-8 \left(w^6+3 w^4-w^2+2\right)\right)\right. \\
		&\left.+\text{ks}^6 \left(4 v^6+v^4 \left(48-206 w^2\right)+v^2 \left(-206 w^4+254 w^2+32\right)+4 \left(w^6+12 w^4+8 w^2+8\right)\right)\right. \\
		&\left.-3 \text{ks}^4 \left(8 v^4+v^2 \left(37 w^2+24\right)+8 w^2 \left(w^2+3\right)\right)+4 \text{ks}^2 \left(9 v^2+9 w^2-8\right)+16\right) \ ,
	\end{aligned}
\end{equation}

\begin{equation}
	\begin{aligned}
		&\mathbb{P}^{22}_3=\mathbb{P}^{22} \left( \textbf{k}, \textbf{p}, \textbf{p}-\textbf{q} -	\textbf{k}, \textbf{q}, -\textbf{q} \right) +\mathbb{P}^{22} \left( \textbf{k}, \textbf{p}, \textbf{p}-\textbf{q} - \textbf{k},	\textbf{q}, \textbf{p}- \textbf{k}   \right)\\
		&=-\frac{k^2}{512 \text{ks}^{10} v^2 \left(\text{ks}^2 \left(v^2+w^2-1\right)-3\right)} \left(\text{ks}^{12} v^2 \left(v^{10}+3 v^8 \left(w^2-1\right)+v^6 \left(-37 w^4+36 w^2+1\right)\right.\right. \\
		&\left.\left.-v^4 \left(w^2-1\right)^2 \left(79 w^2-3\right)+v^2 \left(-32 w^8+98 w^6-109 w^4+44 w^2-2\right)+w^2 \left(8 w^8-32 w^6+41 w^4-19 w^2\right.\right.\right. \\
		&\left.\left.\left.+2\right)\right)+\text{ks}^{10} \left(-9 v^{10}+20 v^8+v^6 \left(267 w^4-266 w^2-1\right)+v^4 \left(234 w^6-488 w^4+254 w^2+26\right)+v^2 \left(-48 w^8\right.\right.\right. \\
		&\left.\left.\left.+134 w^6-113 w^4+66 w^2-36\right)+4 w^2 \left(w^4-3 w^2+2\right)\right)+\text{ks}^8 \left(27 v^8-3 v^6 \left(27 w^2+13\right)+v^4 \left(-459 w^4\right.\right.\right. \\
		&\left.\left.\left.+552 w^2-66\right)+v^2 \left(81 w^6-45 w^4-66 w^2-48\right)-8 w^6+12 w^4+8 w^2+16\right)+\text{ks}^6 \left(-27 v^6 \right.\right. \\
		&\left.\left.+10 v^4 \left(11 w^2+3\right)+v^2 \left(-79 w^4-170 w^2+168\right)+4 \left(w^6+3 w^4-10 w^2-8\right)\right)\right. \\
		&\left.+12 \text{ks}^4 \left(v^4+2 v^2 \left(7 w^2-2\right)-w^2 \left(w^2-2\right)\right)-4 \text{ks}^2 \left(9 v^2-8\right)-16\right)\ ,
	\end{aligned}
\end{equation}

\begin{equation}
	\begin{aligned}
		\mathbb{P}^{33}_1=\mathbb{P}^{33}	\left( \textbf{k}, \textbf{p}, - \textbf{p}, \textbf{q}, - \textbf{q}	\right) +\mathbb{P}^{33} \left( \textbf{k}, \textbf{p}, -	\textbf{p}, \textbf{q}, \textbf{q} - \textbf{p} \right) = 0\ ,
	\end{aligned}
\end{equation}

\begin{equation}
	\begin{aligned}
		\mathbb{P}^{33}_2=\mathbb{P}^{33} \left( \textbf{k}, \textbf{p}, \textbf{q} -	\textbf{k}, \textbf{q}, \textbf{q} - \textbf{p} \right) +\mathbb{P}^{33} \left( \textbf{k}, \textbf{p}, \textbf{q} - \textbf{k},	\textbf{q}, - \textbf{k} + \textbf{p} \right)= 0\ ,
	\end{aligned}
\end{equation}

\begin{equation}
	\begin{aligned}
		\mathbb{P}^{33}_3=\mathbb{P}^{33} \left( \textbf{k}, \textbf{p}, \textbf{p}-\textbf{q} -	\textbf{k}, \textbf{q}, -\textbf{q} \right) +\mathbb{P}^{33} \left( \textbf{k}, \textbf{p}, \textbf{p}-\textbf{q} - \textbf{k},	\textbf{q}, \textbf{p}- \textbf{k}   \right)=0 \ ,
	\end{aligned}
\end{equation}

\begin{equation}
	\begin{aligned}
		&\mathbb{P}^{44}_1=\mathbb{P}^{44}	\left( \textbf{k}, \textbf{p}, - \textbf{p}, \textbf{q}, - \textbf{q}	\right) +\mathbb{P}^{44} \left( \textbf{k}, \textbf{p}, -	\textbf{p}, \textbf{q}, \textbf{q} - \textbf{p} \right) \\
		&=-\frac{k^2 \left(\text{ks}^4 \left(v^2-1\right)^2-2 \text{ks}^2 \left(v^2+1\right)+1\right) \left(\text{ks}^2 \left(v^2+1\right)-1\right)^2}{8 \text{ks}^8} \ ,
	\end{aligned}
\end{equation}

\begin{equation}
	\begin{aligned}
		&\mathbb{P}^{44}_2=\mathbb{P}^{44} \left( \textbf{k}, \textbf{p}, \textbf{q} -	\textbf{k}, \textbf{q}, \textbf{q} - \textbf{p} \right) +\mathbb{P}^{44} \left( \textbf{k}, \textbf{p}, \textbf{q} - \textbf{k},	\textbf{q}, - \textbf{k} + \textbf{p} \right)\\
		&=-\frac{k^2 \left(\text{ks}^2 \left(v^2+1\right)-1\right) \left(\text{ks}^2 \left(w^2+1\right)-1\right) \left(\text{ks}^4 \left(v^2 \left(w^2+1\right)+w^2-1\right)-\text{ks}^2 \left(v^2+w^2\right)+1\right)}{8 \text{ks}^8} \ ,
	\end{aligned}
\end{equation}

\begin{equation}
	\begin{aligned}
		&\mathbb{P}^{44}_3=\mathbb{P}^{44} \left( \textbf{k}, \textbf{p}, \textbf{p}-\textbf{q} -	\textbf{k}, \textbf{q}, -\textbf{q} \right) +\mathbb{P}^{44} \left( \textbf{k}, \textbf{p}, \textbf{p}-\textbf{q} - \textbf{k},	\textbf{q}, \textbf{p}- \textbf{k}   \right)\\
		&=-\frac{k^2 \left(\text{ks}^2 \left(v^2+1\right)-1\right) \left(\text{ks}^2 \left(v^2+w^2-2\right)-2\right) \left(\text{ks}^4 \left(v^4+v^2 \left(w^2-1\right)+w^2\right)-\text{ks}^2 \left(3 v^2+w^2+2\right)+2\right)}{8 \text{ks}^8} \ ,
	\end{aligned}
\end{equation}

\begin{equation}
	\begin{aligned}
		&\mathbb{P}^{12}_1=\mathbb{P}^{12}	\left( \textbf{k}, \textbf{p}, - \textbf{p}, \textbf{q}, - \textbf{q}	\right) +\mathbb{P}^{12} \left( \textbf{k}, \textbf{p}, -	\textbf{p}, \textbf{q}, \textbf{q} - \textbf{p} \right)\\
		&=\frac{k^2}{64 \text{ks}^{10} v^2} \left(\text{ks}^2 \left(v^2+w^2\right)-2\right) \left(\text{ks}^8 v^2 \left(v^4 \left(w^2+1\right)+v^2 \left(8 w^4-6 w^2\right)+8 w^6-16 w^4+9 w^2-1\right) \right. \\
		&\left.-\text{ks}^6 \left(v^6+2 v^4 \left(7 w^2+3\right)+v^2 \left(32 w^4-22 w^2-1\right)-4 w^2+4\right)+\text{ks}^4 \left(6 v^4+v^2 \left(33 w^2+9\right)-8 w^2+4\right)\right. \\
		&\left.+\text{ks}^2 \left(-9 v^2+4 w^2+4\right)-4\right) \ ,
	\end{aligned}
\end{equation}

\begin{equation}
	\begin{aligned}
		&\mathbb{P}^{12}_2=\mathbb{P}^{12} \left( \textbf{k}, \textbf{p}, \textbf{q} -	\textbf{k}, \textbf{q}, \textbf{q} - \textbf{p} \right) +\mathbb{P}^{12} \left( \textbf{k}, \textbf{p}, \textbf{q} - \textbf{k},	\textbf{q}, - \textbf{k} + \textbf{p} \right)\\
		&=\frac{k^2}{64 \text{ks}^{10} w^2} \left(\text{ks}^2 \left(w^2+1\right)-1\right) \left(\text{ks}^2 \left(v^2+w^2\right)-2\right) \left(\text{ks}^6 w^2 \left(8 v^4+8 v^2 \left(w^2-1\right)+\left(w^2-1\right)^2\right)\right. \\
		&\left.+\text{ks}^4 \left(\left(6-24 v^2\right) w^2-6 w^4+4\right)+\text{ks}^2 \left(9 w^2-8\right)+4\right) \ ,
	\end{aligned}
\end{equation}

\begin{equation}
	\begin{aligned}
		&\mathbb{P}^{12}_3=\mathbb{P}^{12} \left( \textbf{k}, \textbf{p}, \textbf{p}-\textbf{q} -	\textbf{k}, \textbf{q}, -\textbf{q} \right) +\mathbb{P}^{12} \left( \textbf{k}, \textbf{p}, \textbf{p}-\textbf{q} - \textbf{k},	\textbf{q}, \textbf{p}- \textbf{k}   \right)\\
		&=\frac{k^2}{64 \text{ks}^8 \left(\text{ks}^2 \left(v^2+w^2-1\right)-3\right)} \left(\text{ks}^2 \left(v^2+w^2\right)-2\right) \left(\text{ks}^8 \left(v^6 \left(w^2+1\right)+v^4 \left(-5 w^4+6 w^2-3\right)\right. \right.\\
		&\left.\left.+v^2 \left(-5 w^6+9 w^4-6 w^2+2\right)+w^2 \left(w^2-2\right) \left(w^2-1\right)^2\right)+\text{ks}^6 \left(-v^6+2 v^4 w^2+v^2 \left(23 w^4-24 w^2+4\right)\right.\right. \\
		&\left.\left.-4 w^6+8 w^4-8 w^2+4\right)+\text{ks}^4 \left(3 v^4-6 v^2 \left(3 w^2+1\right)+3 w^4+14 w^2-4\right)-4 \text{ks}^2 \left(w^2+1\right)+4\right) \ ,
	\end{aligned}
\end{equation}

\begin{equation}
	\begin{aligned}
		&\mathbb{P}^{21}_1=\mathbb{P}^{21}	\left( \textbf{k}, \textbf{p}, - \textbf{p}, \textbf{q}, - \textbf{q}	\right) +\mathbb{P}^{21} \left( \textbf{k}, \textbf{p}, -	\textbf{p}, \textbf{q}, \textbf{q} - \textbf{p} \right) \\
		&=\frac{k^2}{64 \text{ks}^{10} v^2} \left(\text{ks}^{10} v^2 w^2 \left(v^6+v^4 \left(9 w^2-7\right)+v^2 \left(16 w^4-22 w^2+7\right)+8 w^6-16 w^4+9 w^2-1\right)\right. \\
		&\left.-\text{ks}^8 \left(v^8+17 v^6 w^2+v^4 \left(62 w^4-38 w^2-3\right)+v^2 \left(48 w^6-54 w^4+5 w^2+6\right)-4 w^2 \left(w^2-1\right)\right)\right. \\
		&\left.+\text{ks}^6 \left(8 v^6+v^4 \left(67 w^2+2\right)+v^2 \left(97 w^4-51 w^2-6\right)-4 \left(2 w^4+w^2-2\right)\right)-\text{ks}^4 \left(21 v^4+v^2 \left(71 w^2+2\right)\right.\right. \\
		&\left.\left.-4 \left(w^4+5 w^2-2\right)\right)+2 \text{ks}^2 \left(7 v^2-6 w^2-4\right)+8\right)\ ,
	\end{aligned}
\end{equation}

\begin{equation}
	\begin{aligned}
		&\mathbb{P}^{21}_2=\mathbb{P}^{21} \left( \textbf{k}, \textbf{p}, \textbf{q} -	\textbf{k}, \textbf{q}, \textbf{q} - \textbf{p} \right) +\mathbb{P}^{21} \left( \textbf{k}, \textbf{p}, \textbf{q} - \textbf{k},	\textbf{q}, - \textbf{k} + \textbf{p} \right)\\
		&=\frac{k^2}{64 \text{ks}^{10} v^2} \left(\text{ks}^{10} v^4 \left(v^6+v^4 \left(9 w^2-1\right)+v^2 \left(16 w^4-8 w^2-1\right)+8 w^6-8 w^4-w^2+1\right)+\text{ks}^8 v^2 \left(-9 v^6\right.\right. \\
		&\left.\left.+v^4 \left(4-55 w^2\right)+v^2 \left(-56 w^4+22 w^2+13\right)-8 w^6+8 w^4+11 w^2-4\right)+\text{ks}^6 \left(29 v^6+v^4 \left(103 w^2-11\right)\right.\right. \\
		&\left.\left.+v^2 \left(40 w^4-22 w^2-26\right)-4 w^2\right)+\text{ks}^4 \left(-35 v^4+v^2 \left(24-53 w^2\right)+8 \left(w^2+1\right)\right)\right. \\
		&\left.+2 \text{ks}^2 \left(3 v^2-2 \left(w^2+4\right)\right)+8\right) \ ,
	\end{aligned}
\end{equation}

\begin{equation}
	\begin{aligned}
		&\mathbb{P}^{21}_3=\mathbb{P}^{21} \left( \textbf{k}, \textbf{p}, \textbf{p}-\textbf{q} -	\textbf{k}, \textbf{q}, -\textbf{q} \right) +\mathbb{P}^{21} \left( \textbf{k}, \textbf{p}, \textbf{p}-\textbf{q} - \textbf{k},	\textbf{q}, \textbf{p}- \textbf{k}   \right)\\
		&=-\frac{k^2}{64 \text{ks}^{10} v^2} \left(\text{ks}^{10} v^4 w^2 \left(v^4+8 v^2 \left(w^2-1\right)+8 w^4-16 w^2+7\right)+\text{ks}^8 v^2 \left(-v^6+v^4 \left(1-15 w^2\right)\right. \right.\\
		&\left.\left.+v^2 \left(-40 w^4+38 w^2+3\right)-8 w^6+16 w^4+3 w^2-7\right)+\text{ks}^6 \left(7 v^6+v^4 \left(47 w^2-4\right)+v^2 \left(32 w^4-38 w^2-5\right)\right.\right. \\
		&\left.\left.-4 w^2+4\right)+\text{ks}^4 \left(-15 v^4+v^2 \left(7-29 w^2\right)+8 w^2-4\right)+\text{ks}^2 \left(5 v^2-4 \left(w^2+1\right)\right)+4\right)  \ ,
	\end{aligned}
\end{equation}

\begin{equation}
	\begin{aligned}
		\mathbb{P}^{13}_1=\mathbb{P}^{13}	\left( \textbf{k}, \textbf{p}, - \textbf{p}, \textbf{q}, - \textbf{q}	\right) +\mathbb{P}^{13} \left( \textbf{k}, \textbf{p}, -	\textbf{p}, \textbf{q}, \textbf{q} - \textbf{p} \right) = 0\ ,
	\end{aligned}
\end{equation}

\begin{equation}
	\begin{aligned}
		\mathbb{P}^{13}_2=\mathbb{P}^{13} \left( \textbf{k}, \textbf{p}, \textbf{q} -	\textbf{k}, \textbf{q}, \textbf{q} - \textbf{p} \right) +\mathbb{P}^{13} \left( \textbf{k}, \textbf{p}, \textbf{q} - \textbf{k},	\textbf{q}, - \textbf{k} + \textbf{p} \right)= 0\ ,
	\end{aligned}
\end{equation}

\begin{equation}
	\begin{aligned}
		\mathbb{P}^{13}_3=\mathbb{P}^{13} \left( \textbf{k}, \textbf{p}, \textbf{p}-\textbf{q} -	\textbf{k}, \textbf{q}, -\textbf{q} \right) +\mathbb{P}^{13} \left( \textbf{k}, \textbf{p}, \textbf{p}-\textbf{q} - \textbf{k},	\textbf{q}, \textbf{p}- \textbf{k}   \right)=0 \ ,
	\end{aligned}
\end{equation}

\begin{equation}
	\begin{aligned}
		&\mathbb{P}^{31}_1=\mathbb{P}^{31}	\left( \textbf{k}, \textbf{p}, - \textbf{p}, \textbf{q}, - \textbf{q}	\right) +\mathbb{P}^{31} \left( \textbf{k}, \textbf{p}, -	\textbf{p}, \textbf{q}, \textbf{q} - \textbf{p} \right) \\
		&= -\frac{k^2 \left(\text{ks}^2 \left(v^2+1\right)-1\right) \left(\text{ks}^2 \left(v^2+2 w^2-1\right)-3\right) \left(\text{ks}^4 w^2 \left(v^2+w^2-1\right)-\text{ks}^2 \left(v^2+3 w^2\right)+2\right)}{16 \text{ks}^8}\ ,
	\end{aligned}
\end{equation}

\begin{equation}
	\begin{aligned}
		&\mathbb{P}^{31}_2=\mathbb{P}^{31} \left( \textbf{k}, \textbf{p}, \textbf{q} -	\textbf{k}, \textbf{q}, \textbf{q} - \textbf{p} \right) +\mathbb{P}^{31} \left( \textbf{k}, \textbf{p}, \textbf{q} - \textbf{k},	\textbf{q}, - \textbf{k} + \textbf{p} \right) \\
		&=-\frac{k^2}{16 \text{ks}^8} \left(\text{ks}^8 v^2 \left(v^6+v^4 \left(3 w^2-1\right)+v^2 \left(2 w^4-2 w^2+1\right)+w^2-1\right)-\text{ks}^6 \left(7 v^6+v^4 \left(13 w^2-2\right)\right.\right. \\
		&\left.\left.+v^2 \left(4 w^4+1\right)+2 w^4-w^2-2\right)+\text{ks}^4 \left(17 v^4+v^2 \left(17 w^2+3\right)+2 \left(w^4+3 w^2-1\right)\right)\right. \\
		&\left.-\text{ks}^2 \left(17 v^2+7 w^2+6\right)+6\right) \ ,
	\end{aligned}
\end{equation}

\begin{equation}
	\begin{aligned}
		&\mathbb{P}^{31}_3=\mathbb{P}^{31} \left( \textbf{k}, \textbf{p}, \textbf{p}-\textbf{q} -	\textbf{k}, \textbf{q}, -\textbf{q} \right) +\mathbb{P}^{31} \left( \textbf{k}, \textbf{p}, \textbf{p}-\textbf{q} - \textbf{k},	\textbf{q}, \textbf{p}- \textbf{k}   \right) \\
		&=\frac{k^2}{16 \text{ks}^8} \left(\text{ks}^8 v^2 w^2 \left(v^4+2 v^2 \left(w^2-1\right)-1\right)+\text{ks}^6 \left(-v^6+v^4 \left(1-7 w^2\right)+v^2 \left(-4 w^4+4 w^2+1\right)-2 w^4\right.\right. \\
		&\left.\left.+3 w^2+1\right)+\text{ks}^4 \left(5 v^4+v^2 \left(11 w^2-2\right)+2 w^4+2 w^2-3\right)-\text{ks}^2 \left(7 v^2+5 w^2+1\right)+3\right)  \ ,
	\end{aligned}
\end{equation}

\begin{equation}
	\begin{aligned}
		&\mathbb{P}^{14}_1=\mathbb{P}^{14}	\left( \textbf{k}, \textbf{p}, - \textbf{p}, \textbf{q}, - \textbf{q}	\right) +\mathbb{P}^{14} \left( \textbf{k}, \textbf{p}, -	\textbf{p}, \textbf{q}, \textbf{q} - \textbf{p} \right) \\
		&=\frac{k^2 \left(\text{ks}^2 \left(v^2+1\right)-1\right) \left(\text{ks}^2 \left(v^2+w^2\right)-2\right) \left(\text{ks}^4 \left(v^2 \left(w^2+1\right)+w^2-1\right)-\text{ks}^2 \left(v^2+w^2\right)+1\right)}{8 \text{ks}^8} \ ,
	\end{aligned}
\end{equation}

\begin{equation}
	\begin{aligned}
		&\mathbb{P}^{14}_2=\mathbb{P}^{14} \left( \textbf{k}, \textbf{p}, \textbf{q} -	\textbf{k}, \textbf{q}, \textbf{q} - \textbf{p} \right) +\mathbb{P}^{14} \left( \textbf{k}, \textbf{p}, \textbf{q} - \textbf{k},	\textbf{q}, - \textbf{k} + \textbf{p} \right) \\
		&= \frac{k^2 \left(\text{ks}^4 \left(w^2-1\right)^2-2 \text{ks}^2 \left(w^2+1\right)+1\right) \left(\text{ks}^2 \left(w^2+1\right)-1\right) \left(\text{ks}^2 \left(v^2+w^2\right)-2\right)}{8 \text{ks}^8}\ ,
	\end{aligned}
\end{equation}

\begin{equation}
	\begin{aligned}
		&\mathbb{P}^{14}_3=\mathbb{P}^{14} \left( \textbf{k}, \textbf{p}, \textbf{p}-\textbf{q} -	\textbf{k}, \textbf{q}, -\textbf{q} \right) +\mathbb{P}^{14} \left( \textbf{k}, \textbf{p}, \textbf{p}-\textbf{q} - \textbf{k},	\textbf{q}, \textbf{p}- \textbf{k}   \right) \\
		&=\frac{k^2}{8 \text{ks}^8} \left(\text{ks}^2 \left(v^2+w^2-2\right)-2\right) \left(\text{ks}^2 \left(v^2+w^2\right)-2\right) \left(\text{ks}^4 \left(v^2 \left(w^2+1\right)+w^2 \left(w^2-1\right)\right) \right. \\
		&\left.-\text{ks}^2 \left(v^2+3 w^2+2\right)+2\right)  \ ,
	\end{aligned}
\end{equation}

\begin{equation}
	\begin{aligned}
		&\mathbb{P}^{41}_1=\mathbb{P}^{41}	\left( \textbf{k}, \textbf{p}, - \textbf{p}, \textbf{q}, - \textbf{q}	\right) +\mathbb{P}^{41} \left( \textbf{k}, \textbf{p}, -	\textbf{p}, \textbf{q}, \textbf{q} - \textbf{p} \right) \\
		&=\frac{k^2 \left(\text{ks}^2 \left(v^2+1\right)-1\right)^2 \left(\text{ks}^4 w^2 \left(v^2+w^2-1\right)-\text{ks}^2 \left(v^2+3 w^2-2\right)+2\right)}{8 \text{ks}^8} \ ,
	\end{aligned}
\end{equation}

\begin{equation}
	\begin{aligned}
		&\mathbb{P}^{41}_2=\mathbb{P}^{41} \left( \textbf{k}, \textbf{p}, \textbf{q} -	\textbf{k}, \textbf{q}, \textbf{q} - \textbf{p} \right) +\mathbb{P}^{41} \left( \textbf{k}, \textbf{p}, \textbf{q} - \textbf{k},	\textbf{q}, - \textbf{k} + \textbf{p} \right) \\
		&=\frac{k^2}{8 \text{ks}^8} \left(\text{ks}^2 \left(v^2+1\right)-1\right) \left(\text{ks}^6 v^2 \left(v^2-1\right) \left(v^2+w^2-1\right)-\text{ks}^4 \left(4 v^4+2 v^2 \left(w^2-1\right)+w^2\right)\right. \\
		&\left.+\text{ks}^2 \left(5 v^2+w^2+2\right)-2\right)  \ ,
	\end{aligned}
\end{equation}

\begin{equation}
	\begin{aligned}
		&\mathbb{P}^{41}_3=\mathbb{P}^{41} \left( \textbf{k}, \textbf{p}, \textbf{p}-\textbf{q} -	\textbf{k}, \textbf{q}, -\textbf{q} \right) +\mathbb{P}^{41} \left( \textbf{k}, \textbf{p}, \textbf{p}-\textbf{q} - \textbf{k},	\textbf{q}, \textbf{p}- \textbf{k}   \right) \\
		&=-\frac{k^2 \left(\text{ks}^2 \left(v^2+1\right)-1\right) \left(\text{ks}^6 v^2 \left(v^2-1\right) w^2-\text{ks}^4 \left(v^4+2 v^2 \left(w^2-1\right)+w^2-1\right)+\text{ks}^2 \left(2 v^2+w^2\right)-1\right)}{8 \text{ks}^8} \ ,
	\end{aligned}
\end{equation}

\begin{equation}
	\begin{aligned}
		&\mathbb{P}^{23}_1=-\mathbb{P}^{23}	\left( \textbf{k}, \textbf{p}, - \textbf{p}, \textbf{q}, - \textbf{q}	\right) +\mathbb{P}^{23} \left( \textbf{k}, \textbf{p}, -	\textbf{p}, \textbf{q}, \textbf{q} - \textbf{p} \right) =0 \ ,
	\end{aligned}
\end{equation}

\begin{equation}
	\begin{aligned}
		&\mathbb{P}^{23}_2=\mathbb{P}^{23} \left( \textbf{k}, \textbf{p}, \textbf{q} -	\textbf{k}, \textbf{q}, \textbf{q} - \textbf{p} \right) +\mathbb{P}^{23} \left( \textbf{k}, \textbf{p}, \textbf{q} - \textbf{k},	\textbf{q}, - \textbf{k} + \textbf{p} \right) = 0\ ,
	\end{aligned}
\end{equation}

\begin{equation}
	\begin{aligned}
		&\mathbb{P}^{23}_3=\mathbb{P}^{23} \left( \textbf{k}, \textbf{p}, \textbf{p}-\textbf{q} -	\textbf{k}, \textbf{q}, -\textbf{q} \right) +\mathbb{P}^{23} \left( \textbf{k}, \textbf{p}, \textbf{p}-\textbf{q} - \textbf{k},	\textbf{q}, \textbf{p}- \textbf{k}   \right) =0 \ ,
	\end{aligned}
\end{equation}

\begin{equation}
	\begin{aligned}
		&\mathbb{P}^{32}_1=\mathbb{P}^{32}	\left( \textbf{k}, \textbf{p}, - \textbf{p}, \textbf{q}, - \textbf{q}	\right) +\mathbb{P}^{32} \left( \textbf{k}, \textbf{p}, -	\textbf{p}, \textbf{q}, \textbf{q} - \textbf{p} \right) \\
		&=\frac{k^2}{128 \text{ks}^{10} v^2} \left(\text{ks}^2 \left(v^2+1\right)-1\right) \left(\text{ks}^2 \left(v^2+2 w^2-1\right)-3\right) \left(\text{ks}^6 v^2 \left(v^4+8 v^2 w^2+8 w^4-8 w^2+1\right)\right. \\
		&\left.-2 \text{ks}^4 \left(3 v^4+v^2 \left(12 w^2+1\right)-2\right)+\text{ks}^2 \left(9 v^2-8\right)+4\right)\ ,
	\end{aligned}
\end{equation}

\begin{equation}
	\begin{aligned}
		&\mathbb{P}^{32}_2=\mathbb{P}^{32} \left( \textbf{k}, \textbf{p}, \textbf{q} -	\textbf{k}, \textbf{q}, \textbf{q} - \textbf{p} \right) +\mathbb{P}^{32} \left( \textbf{k}, \textbf{p}, \textbf{q} - \textbf{k},	\textbf{q}, - \textbf{k} + \textbf{p} \right) \\
		&=\frac{k^2}{128 \text{ks}^{10} w^2} \left(\text{ks}^{10} w^2 \left(8 v^8+8 v^6 \left(3 w^2-1\right)+v^4 \left(17 w^4-14 w^2+9\right)+2 v^2 \left(w^6+w^4+4 w^2-4\right)+2 w^6 \right.\right. \\
		&\left.\left.-w^4-2 w^2+1\right)-2 \text{ks}^8 \left(28 v^6 w^2+v^4 \left(51 w^4-11 w^2-2\right)+v^2 \left(16 w^4+w^2+6\right) w^2+w^8+7 w^6-2 w^4\right.\right. \\
		&\left.\left.-3 w^2-2\right)+\text{ks}^6 \left(v^4 \left(129 w^2-8\right)+2 v^2 \left(57 w^4-4 w^2-8\right)+15 w^6+34 w^4-12 w^2-8\right)\right. \\
		&\left.+2 \text{ks}^4 \left(2 v^4+v^2 \left(16-50 w^2\right)-18 w^4-7 w^2+8\right)+\text{ks}^2 \left(-16 v^2+19 w^2-24\right)+12\right)\ ,
	\end{aligned}
\end{equation}

\begin{equation}
	\begin{aligned}
		&\mathbb{P}^{32}_3=\mathbb{P}^{32} \left( \textbf{k}, \textbf{p}, \textbf{p}-\textbf{q} -	\textbf{k}, \textbf{q}, -\textbf{q} \right) +\mathbb{P}^{32} \left( \textbf{k}, \textbf{p}, \textbf{p}-\textbf{q} - \textbf{k},	\textbf{q}, \textbf{p}- \textbf{k}   \right) \\
		&=\frac{k^2}{128 \text{ks}^8 \left(\text{ks}^2 \left(v^2+w^2-1\right)-3\right)} \left(\text{ks}^{10} \left(v^{10}-3 v^8 \left(w^2+1\right)+v^6 \left(-15 w^4+12 w^2+3\right)-v^4 \left(9 w^6-19 w^4 \right.\right.\right. \\
		&\left.\left.\left.+w^2+3\right)+v^2 \left(2 w^8-2 w^6+5 w^4-8 w^2+2\right)+w^2 \left(2 w^6-7 w^4+7 w^2-2\right)\right)-\text{ks}^8 \left(7 v^8-2 v^6 \left(15 w^2+7\right)\right.\right. \\
		&\left.\left.+v^4 \left(-63 w^4+62 w^2+11\right)+2 v^2 \left(7 w^4+9 w^2-7\right)+2 w^8+2 w^6-13 w^4+6 w^2-4\right)+\text{ks}^6 \left(15 v^6\right.\right. \\
		&\left.\left.-v^4 \left(81 w^2+5\right)+v^2 \left(-39 w^4+76 w^2-2\right)+9 w^6-11 w^4+10 w^2-16\right)-\text{ks}^4 \left(13 v^4+v^2 \left(30-46 w^2\right)\right.\right. \\
		&\left.\left.+9 w^4+10 w^2-8\right)+8 \text{ks}^2 \left(2 v^2+w^2+2\right)-12\right) \ ,
	\end{aligned}
\end{equation}

\begin{equation}
	\begin{aligned}
		&\mathbb{P}^{24}_1=\mathbb{P}^{24}	\left( \textbf{k}, \textbf{p}, - \textbf{p}, \textbf{q}, - \textbf{q}	\right) +\mathbb{P}^{24} \left( \textbf{k}, \textbf{p}, -	\textbf{p}, \textbf{q}, \textbf{q} - \textbf{p} \right) \\
		&=-\frac{k^2}{64 \text{ks}^{10} v^2} \left(\text{ks}^2 \left(v^2+1\right)-1\right)^2 \left(\text{ks}^6 v^2 \left(v^4+v^2 \left(8 w^2-2\right)+8 w^4-8 w^2+1\right)\right.\\
		&\left.+\text{ks}^4 \left(-6 v^4+v^2 \left(6-24 w^2\right)+4\right)+\text{ks}^2 \left(9 v^2-8\right)+4\right) \ ,
	\end{aligned}
\end{equation}

\begin{equation}
	\begin{aligned}
		&\mathbb{P}^{24}_2=\mathbb{P}^{24} \left( \textbf{k}, \textbf{p}, \textbf{q} -	\textbf{k}, \textbf{q}, \textbf{q} - \textbf{p} \right) +\mathbb{P}^{24} \left( \textbf{k}, \textbf{p}, \textbf{q} - \textbf{k},	\textbf{q}, - \textbf{k} + \textbf{p} \right) \\
		&=-\frac{k^2}{64 \text{ks}^{10} v^2} \left(\text{ks}^2 \left(w^2+1\right)-1\right) \left(\text{ks}^8 v^2 \left(v^4 \left(w^2+1\right)+v^2 \left(8 w^4-6 w^2\right)+8 w^6-16 w^4+9 w^2-1\right)\right. \\
		&\left.-\text{ks}^6 \left(v^6+2 v^4 \left(7 w^2+3\right)+v^2 \left(32 w^4-22 w^2-1\right)-4 w^2+4\right)\right. \\
		&\left.+\text{ks}^4 \left(6 v^4+v^2 \left(33 w^2+9\right)-8 w^2+4\right)+\text{ks}^2 \left(-9 v^2+4 w^2+4\right)-4\right) \ ,
	\end{aligned}
\end{equation}

\begin{equation}
	\begin{aligned}
		&\mathbb{P}^{24}_3=\mathbb{P}^{24} \left( \textbf{k}, \textbf{p}, \textbf{p}-\textbf{q} -	\textbf{k}, \textbf{q}, -\textbf{q} \right) +\mathbb{P}^{24} \left( \textbf{k}, \textbf{p}, \textbf{p}-\textbf{q} - \textbf{k},	\textbf{q}, \textbf{p}- \textbf{k}   \right) \\
		&=-\frac{k^2}{64 \text{ks}^{10} v^2} \left(\text{ks}^2 \left(v^2+w^2-2\right)-2\right) \left(\text{ks}^8 v^2 \left(v^6+9 v^4 w^2+v^2 \left(16 w^4-6 w^2-1\right)+8 w^6-8 w^4+w^2\right)\right. \\
		&\left.-2 \text{ks}^6 \left(4 v^6+v^4 \left(23 w^2+2\right)+v^2 \left(20 w^4-3 w^2-5\right)-2 w^2\right)+\text{ks}^4 \left(21 v^4+v^2 \left(57 w^2+4\right)-8 \left(w^2+1\right)\right)\right. \\
		&\left.+\text{ks}^2 \left(4 \left(w^2+4\right)-14 v^2\right)-8\right) \ ,
	\end{aligned}
\end{equation}

\begin{equation}
	\begin{aligned}
		&\mathbb{P}^{42}_1=\mathbb{P}^{42}	\left( \textbf{k}, \textbf{p}, - \textbf{p}, \textbf{q}, - \textbf{q}	\right) +\mathbb{P}^{42} \left( \textbf{k}, \textbf{p}, -	\textbf{p}, \textbf{q}, \textbf{q} - \textbf{p} \right) \\
		&=-\frac{k^2}{64 \text{ks}^{10} v^2} \left(\text{ks}^2 \left(v^2+1\right)-1\right)^2 \left(\text{ks}^6 v^2 \left(v^4+v^2 \left(8 w^2-2\right)+8 w^4-8 w^2+1\right)\right. \\
		&\left.+\text{ks}^4 \left(-6 v^4+v^2 \left(6-24 w^2\right)+4\right)+\text{ks}^2 \left(9 v^2-8\right)+4\right) \ ,
	\end{aligned}
\end{equation}

\begin{equation}
	\begin{aligned}
		&\mathbb{P}^{42}_2=\mathbb{P}^{42} \left( \textbf{k}, \textbf{p}, \textbf{q} -	\textbf{k}, \textbf{q}, \textbf{q} - \textbf{p} \right) +\mathbb{P}^{42} \left( \textbf{k}, \textbf{p}, \textbf{q} - \textbf{k},	\textbf{q}, - \textbf{k} + \textbf{p} \right) \\
		&=-\frac{k^2}{64 \text{ks}^{10} w^2} \left(\text{ks}^2 \left(v^2+1\right)-1\right) \left(\text{ks}^8 w^2 \left(8 v^6+8 v^4 \left(w^2-2\right)+v^2 \left(w^2-3\right)^2+w^4-1\right)\right. \\
		&\left.-\text{ks}^6 \left(32 v^4 w^2+2 v^2 \left(7 w^4-11 w^2-2\right)+w^6+6 w^4-w^2+4\right)+\text{ks}^4 \left(v^2 \left(33 w^2-8\right)+6 w^4+9 w^2+4\right)\right. \\
		&\left.+\text{ks}^2 \left(4 v^2-9 w^2+4\right)-4\right) \ ,
	\end{aligned}
\end{equation}

\begin{equation}
	\begin{aligned}
		&\mathbb{P}^{42}_3=\mathbb{P}^{42} \left( \textbf{k}, \textbf{p}, \textbf{p}-\textbf{q} -	\textbf{k}, \textbf{q}, -\textbf{q} \right) +\mathbb{P}^{42} \left( \textbf{k}, \textbf{p}, \textbf{p}-\textbf{q} - \textbf{k},	\textbf{q}, \textbf{p}- \textbf{k}   \right) \\
		&=-\frac{k^2}{64 \text{ks}^8 \left(\text{ks}^2 \left(v^2+w^2-1\right)-3\right)} \left(\text{ks}^2 \left(v^2+1\right)-1\right) \left(\text{ks}^8 \left(v^8-v^6 \left(5 w^2+4\right)+v^4 \left(-5 w^4+9 w^2+5\right)\right.\right. \\
		&\left.\left.+v^2 \left(w^6+6 w^4-6 w^2-2\right)+w^2 \left(w^4-3 w^2+2\right)\right)+\text{ks}^6 \left(-4 v^6+v^4 \left(23 w^2+8\right)+2 v^2 \left(w^4-12 w^2-4\right)\right.\right. \\
		&\left.\left.-w^6+4 w^2+4\right)+\text{ks}^4 \left(3 v^4-2 v^2 \left(9 w^2-7\right)+3 w^4-6 w^2-4\right)-4 \text{ks}^2 \left(v^2+1\right)+4\right) \ ,
	\end{aligned}
\end{equation}

\begin{equation}
	\begin{aligned}
		&\mathbb{P}^{34}_1=\mathbb{P}^{34}	\left( \textbf{k}, \textbf{p}, - \textbf{p}, \textbf{q}, - \textbf{q}	\right) +\mathbb{P}^{34} \left( \textbf{k}, \textbf{p}, -	\textbf{p}, \textbf{q}, \textbf{q} - \textbf{p} \right) \\
		&=\frac{k^2 \left(\text{ks}^2 \left(v^2+1\right)-1\right) \left(\text{ks}^4 \left(v^4+1\right)-2 \text{ks}^2 \left(v^2+1\right)+1\right) \left(\text{ks}^2 \left(v^2+2 w^2-1\right)-3\right)}{16 \text{ks}^8} \ ,
	\end{aligned}
\end{equation}

\begin{equation}
	\begin{aligned}
		&\mathbb{P}^{34}_2=\mathbb{P}^{34} \left( \textbf{k}, \textbf{p}, \textbf{q} -	\textbf{k}, \textbf{q}, \textbf{q} - \textbf{p} \right) +\mathbb{P}^{34} \left( \textbf{k}, \textbf{p}, \textbf{q} - \textbf{k},	\textbf{q}, - \textbf{k} + \textbf{p} \right) \\
		&=\frac{k^2}{16 \text{ks}^8} \left(\text{ks}^2 \left(w^2+1\right)-1\right) \left(\text{ks}^6 \left(v^4 \left(w^2+1\right)+2 v^2 w^4+2 w^4-3 w^2+1\right) \right. \\
		&\left.-\text{ks}^4 \left(v^4+v^2 \left(6 w^2+4\right)+2 w^4+2 w^2+1\right)+\text{ks}^2 \left(4 v^2+5 w^2+3\right)-3\right) \ ,
	\end{aligned}
\end{equation}

\begin{equation}
	\begin{aligned}
		&\mathbb{P}^{34}_3=\mathbb{P}^{34} \left( \textbf{k}, \textbf{p}, \textbf{p}-\textbf{q} -	\textbf{k}, \textbf{q}, -\textbf{q} \right) +\mathbb{P}^{34} \left( \textbf{k}, \textbf{p}, \textbf{p}-\textbf{q} - \textbf{k},	\textbf{q}, \textbf{p}- \textbf{k}   \right) \\
		&=\frac{k^2}{16 \text{ks}^8} \left(\text{ks}^2 \left(v^2+w^2-2\right)-2\right) \left(\text{ks}^6 \left(v^6+3 v^4 w^2+2 v^2 w^4+v^2+2 w^4-w^2\right)\right. \\
		&\left.-2 \text{ks}^4 \left(3 v^4+v^2 \left(5 w^2+2\right)+w^4+3 w^2+1\right)+\text{ks}^2 \left(11 v^2+7 w^2+8\right)-6\right) \ ,
	\end{aligned}
\end{equation}

\begin{equation}
	\begin{aligned}
		&\mathbb{P}^{43}_1=\mathbb{P}^{43}	\left( \textbf{k}, \textbf{p}, - \textbf{p}, \textbf{q}, - \textbf{q}	\right) +\mathbb{P}^{43} \left( \textbf{k}, \textbf{p}, -	\textbf{p}, \textbf{q}, \textbf{q} - \textbf{p} \right) =0 \ ,
	\end{aligned}
\end{equation}

\begin{equation}
	\begin{aligned}
		&\mathbb{P}^{43}_2=\mathbb{P}^{43} \left( \textbf{k}, \textbf{p}, \textbf{q} -	\textbf{k}, \textbf{q}, \textbf{q} - \textbf{p} \right) +\mathbb{P}^{43} \left( \textbf{k}, \textbf{p}, \textbf{q} - \textbf{k},	\textbf{q}, - \textbf{k} + \textbf{p} \right) = 0\ ,
	\end{aligned}
\end{equation}

\begin{equation}
	\begin{aligned}
		&\mathbb{P}^{43}_3=\mathbb{P}^{43} \left( \textbf{k}, \textbf{p}, \textbf{p}-\textbf{q} -	\textbf{k}, \textbf{q}, -\textbf{q} \right) +\mathbb{P}^{43} \left( \textbf{k}, \textbf{p}, \textbf{p}-\textbf{q} - \textbf{k},	\textbf{q}, \textbf{p}- \textbf{k}   \right) =0 \ .
	\end{aligned}
\end{equation}

\bibliography{biblio}

\end{document}